# Discovering Structural, Electronic, and Excitonic Properties of Bulk, Nanostructured, and Doped $C_3N_4$ in Diamond- and Graphitic-Like Phases

Da Chen,[#] Pietro Andreozzi,[#] Giulia Frigerio, Daniele Perilli, Paulo Siani, and Cristiana Di Valentin*

**ABSTRACT:** In this systematic density functional theory study, we compare a standard gradient-corrected functional (PBE) with a long-range hybrid functional (HSE06), with and without correction for the dispersion forces, relative to their ability to correctly reproduce structural and electronic properties of different bulk 3D $C_3N_4$ phases, encompassing diamond- and graphitic-like models. Corrugation is found to provide further stabilization to the layered structures with all methods. We observe that the HSE06-D3 method provides results in good agreement with experimental data and with more sophisticated $G_0W_0$ calculations. Based on that, we exploited the method to investigate the nature of the bulk triplet excitons in these $C_3N_4$ structures to evaluate the $S_0$−$T_1$ energy difference, the self-trapping triplet exciton energy, and the photoluminescence emission energy since this is a promising visible-light photocatalyst. Nanostructuring (0D and 2D) is another relevant aspect of these materials in practical applications; therefore, we have considered the effect of single- or multilayer exfoliation or space confinement in nanoparticles. Finally, we also discuss how the introduction of extrinsic dopants (e.g., S atoms) in the nanostructures modifies the atomic and electronic structure.

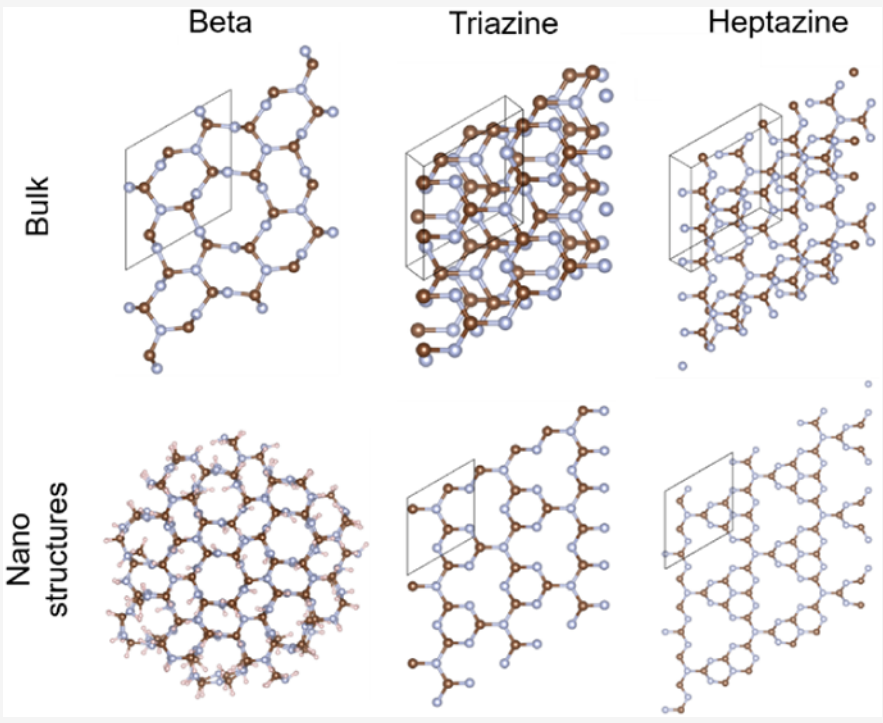

## 1. INTRODUCTION

Carbon nitride has attracted extensive attention due to its distinctive atomic and electronic structures, mechanical properties, and chemical stability.[1−3] Moreover, it is also a promising visible-light-responsive photocatalytic material.[4−6]

Theoretically, carbon nitride is predicted to exist in several potential phases, although not all of them have been successfully synthesized as pure bulk materials under ambient conditions. The most well-known and studied phases fall into two broad categories: rigid three-dimensional (3D) covalent networks (analogous to diamond) or layered graphitic structures.[1−3,7]

The 3D phases are expected to potentially exhibit super hardness, comparable to or even exceeding that of diamond, but they are difficult to synthesize in pure form. Typically, they are prepared as thin films or nanosized crystals (also referred to as C-dots) using high-energy techniques like chemical vapor deposition (CVD) or mechanochemical processing.[8] C-dots are commonly characterized as nanocrystals of the *β*-phase, which is a hexagonal 3D structure analogous to *β*-silicon nitride.

The layered graphitic $C_3N_4$ (g-$C_3N_4$) phase is more experimentally accessible; however, its structure remains unclear since no single-crystal structural analysis could be yet performed.[1,2,9−12] On the basis of crystal structure searching via density functional theory (DFT) calculations, two representative configurations of fully condensed graphitic carbon nitride, i.e., the triazine-based ($C_3N_3$) and the heptazine-based ($C_6N_7$) frameworks, have been proposed, with the latter being more stable than the former by about 30 kJ/mol.[13−15] These ideal hydrogen-free structures are not easily obtained under common synthesis conditions, where linear polymer melon, linked by bridging −N(H)− groups and containing terminal $-NH_2$ groups, is mostly observed.[16,17] The triazine-based g-$C_3N_4$ was successfully obtained in a salt melt.[18,19]

The geometric nature of fully condensed g-$C_3N_4$, whether planar or corrugated, depends strongly on the computational method and dispersion correction applied during structural optimization. Early studies employing the local density approximation (LDA) predicted flat layered structures for the triazine phase, as reported by Teter et al.[14] Subsequent LDA studies confirmed that both triazine and heptazine frameworks tend to form planar sheets in the absence of dispersion corrections.[15,20] Similarly, calculations using the Perdew−Burke−Ernzerhof (PBE) functional also obtained a flat heptazine-based configuration.[21] However, when disper-sion corrections such as D2 or D3 were included, notable differences emerged between the two phases. The triazine phase remained flat even under PBE+D2 optimization,[22] while the heptazine phase exhibited a corrugated structure with PBE +D3 or other vdW corrections.[23−27] These findings indicate that dispersion interactions play a crucial role in stabilizing the layered morphology of g-$C_3N_4$, especially for the heptazine-based system, which tends to adopt a corrugated topology upon inclusion of van der Waals corrections. Similar results are obtained with D3-corrected meta-GGA and hybrid functionals, with large effects of the buckling also on the electronic properties.[28−30]

In this work, using different DFT (standard GGA/PBE and hybrid HSE06) methods, also with the addition of the Grimme correction D3 for the dispersion forces, we investigate the electronic and structural properties of 3D $\beta$-$C_3N_4$ in comparison with fully polymerized layered g-$C_3N_4$ (triazine- and heptazine-based) models, by both imposing or releasing the crystal symmetry constraints with the CRYSTAL23 code. We show that the HSE06-D3 method provides results in quantitatively good agreement with available experimental data (lattice parameters and band gap) and with more sophisticated $G_0W_0$ calculations. The comparative analysis among phases also considers the different characteristics of the photoexcited bulk triplet excitons under irradiation (simulated through spin-constrained calculations) since g-$C_3N_4$ is an excellent photocatalyst. After that, we study the effect of nanostructuring, which in the case of 3D $\beta$-$C_3N_4$ is related to the formation of finite nanocrystals (2 nm), whereas in the case of g-$C_3N_4$ is related to the formation of isolated single/double/triple layers. Finally, we introduce a nonmetal dopant atom in the supercell single layer and nanocrystal models and determine its effect on the electronic properties. The choice to consider an S atom is related to the fact that it was found to be accidentally present in the nanocrystal due to the molecular precursors used for the synthesis.[31] Several studies exist in the literature on the S-doping of graphitic $C_3N_4$ that suggest improved (photo)-catalytic activity with respect to the undoped samples.[32−35]

## 2. COMPUTATIONAL DETAILS

All geometry optimizations and electronic property calculations were performed using DFT as implemented in the CRYSTAL23 code.[36] For bulk (3D), two-dimensional (2D), and functionalized nanoparticle (0D) systems, the Gaussian-type localized m-6-311G(d)_Heyd_2005 basis set was employed. Cutoffs for Coulomb and exchange series in the SCF equations were set to $10^{-7}$ for Coulomb overlap and penetration, $10^{-7}$ for exchange overlap, $10^{-9}$ for exchange pseudo-overlap in direct space, and $10^{-30}$ for exchange pseudo-overlap in reciprocal space. The SCF convergence criterion was $10^{-6}$ a.u. Geometry optimizations were performed with and without symmetry constraints for bulk systems, while for 2D and 0D systems, no symmetry constraints were imposed; atomic forces and displacements were relaxed below $4.5 \times 10^{-4}$ a.u. and $1.8 \times 10^{-3}$ a.u., respectively.

For bulk phases, Perdew−Burke−Ernzerhof (PBE)[37] and Heyd−Scuseria−Ernzerhof (HSE06)[38] functionals were tested, with and without the semiempirical Grimme D3 dispersion correction.[39,40] Accordingly, four levels of theory were considered: PBE, PBE+D3, HSE06, and HSE06+D3. Lattice parameters and atomic positions were fully relaxed. Three bulk systems were considered (space group in parentheses): $\beta$-$C_3N_4$ ($P6_3/m$), triazine-based g-$C_3N_4$ ($R3m$), and heptazine-based g-$C_3N_4$ ($P\bar{6}m2$), with initial structures taken from the Materials Project database.[41] For $\beta$-$C_3N_4$, Monkhorst−Pack[42] k-point meshes of 12 × 12 × 12 and 30 × 30 × 30 were used for geometry optimization and electronic structure calculations, respectively. For triazine- and heptazine-based g-$C_3N_4$, k-point meshes of 3 × 3 × 3 and 20 × 20 × 20 were employed for geometry optimization and electronic structure calculations, respectively.

All systems feature a singlet ground state. Triplet-state calculations were performed to model excitons. Vertical excitations were approximated with a triplet−singlet energy difference at the ground-state geometry ($S_0 \rightarrow T_1$). This choice is imposed by the fact that DFT is a ground-state theory and, therefore, S1 cannot be properly described at this level. T1 is the ground state for the triplet spin configuration and, therefore, can be calculated correctly with DFT methods by imposing a spin constraint.[43] Self-trapping energies correspond to the energy difference of the exciton in trapping versus ground-state geometry ($T_1 \rightarrow T_{1_OPT}$). For these calculations, supercell models of 2 × 2 × 1 were employed for the triazine- and heptazine-based systems, and 2 × 2 × 4 for $\beta$-$C_3N_4$, using a 3 × 3 × 3 k-point mesh in all cases.

For 2D systems, mono-, bi-, and trilayers of triazine-based g-$C_3N_4$, as well as mono- and bilayers of heptazine-based g-$C_3N_4$, were considered. Based on the bulk benchmarks, only the HSE06+D3 functional was employed. Sulfur doping was modeled using 2 × 2 monolayer supercells, resulting in doping concentrations of approximately 8.3 and 4.1 wt % for triazine-based and heptazine-based g-$C_3N_4$, respectively. For both undoped and doped systems, 3 × 3 × 1 and 20 × 20 × 1 k-point meshes were used for geometry optimization and electronic structure calculations, respectively.

Finally, a $\beta$-$C_3N_4$ spherical nanoparticle model was constructed and analyzed in terms of its electronic properties.[44,45] Using the OVITO program,[46] the bulk unit cell was replicated, and a spherical cluster with a radius of 1 nm was extracted. Monocoordinated N atoms and mono- and bicoordinated C atoms were manually removed, retaining only atoms whose coordination differed by at most one relative to the bulk environment, yielding spherical nanoparticles with a diameter of approximately 2 nm. The remaining under-coordinated C and N atoms were saturated with H atoms, resulting in hydrogenated models that preserve the overall $C_3N_4$ stoichiometry. Sulfur doping was also considered by replacing an N−H group with an S atom, resulting in a doping concentration of approximately 0.4 wt %.

All nonperiodic calculations were performed using the HSE06+D3 functional. Total densities of states (DOS) were obtained by convoluting Gaussian functions ($\sigma$ = 0.005 eV) centered at the Kohn−Sham eigenvalues. Projected densities of states (PDOS) were computed from the linear combination of atomic orbitals (LCAO) coefficients by summing the squared coefficients associated with a given atom type and normalizing their contributions. The resulting projections were obtained through Gaussian convolution weighted by the relative contributions. For all DOS plots, the energy zero was set to the vacuum level, corresponding to an electron at infinite distance from the nanoparticle surface.

All ball-and-stick structures have been rendered using the VESTA software.[47]

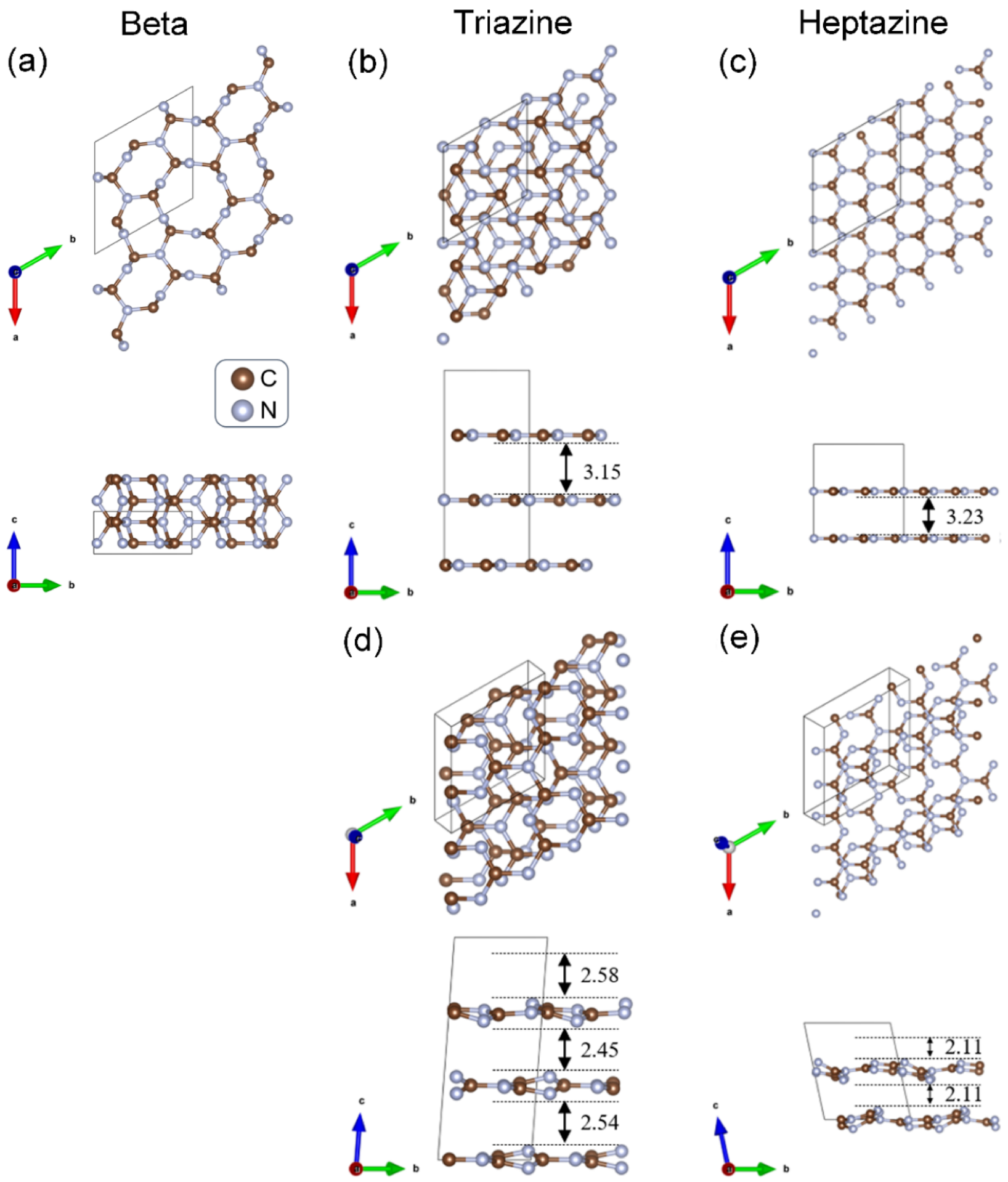

**Figure 1.** Atomic structure of bulk $C_3N_4$ using the HSE06+D3 method: (a) beta phase ($\beta$-$C_3N_4$); (b) triazine-based g-$C_3N_4$ and (c) heptazine-based g-$C_3N_4$ after geometry optimization with (wc) crystal symmetry constraints; (d) triazine-based g-$C_3N_4$ and (e) heptazine-based g-$C_3N_4$ after geometry optimization without (nc) crystal symmetry constraints. The brown and gray spheres represent carbon and nitrogen atoms, respectively.

## 3. RESULTS AND DISCUSSION

### 3.1. Bulk Phases

As we have mentioned in the Introduction, three different bulk phases will be considered in this work. In the top panels of Figure 1, we report the fully relaxed bulk unit cells for (i) the beta phase of $C_3N_4$ ($\beta$-$C_3N_4$), (ii) the triazine-based (tri-$C_3N_4$-nc), and (iii) the heptazine-based (hep-$C_3N_4$-nc) graphitic $C_3N_4$. The "nc" label indicates that no crystal symmetry constraints were imposed during geometry optimization.

$\beta$-$C_3N_4$ possesses a dense, three-dimensional $\beta$-$Si_3N_4$—like framework composed of corner-sharing $CN_4$ tetrahedra, with carbon atoms in the $sp^3$ hybridization state.[14] The space group for $\beta$-$C_3N_4$ is $P6_3/m$, and even if symmetry constraints are removed, it does not change during geometry optimization.

The triazine-based g-$C_3N_4$ is constructed from repeating $C_3N_3$ triazine rings linked by nitrogen atoms, while the heptazine-based form consists of larger $C_6N_7$ heptazine rings connected in a similar manner, with the carbon atoms in both structures being in an $sp^2$ hybridization state.[14,15] In the bottom panels of Figure 1, we also present the optimized structures under crystal symmetry constraints, which means that the geometry optimization was conducted by keeping the starting space group ($R3m$ and $P\overline{6}m2$ from Materials Project[41] for triazine- and heptazine-based g-$C_3N_4$, respectively) fixed, as suggested in previous studies.[21,22] It is evident that in this case, the layers stay flat. However, the most stable configurations of both the triazine- and heptazine-based g-$C_3N_4$ are found to be those obtained without any crystal symmetry constraints, which are characterized by some layer corrugation. The relative phase stability is listed in Table 1. The heptazine-based g-$C_3N_4$ is significantly more stable than the other phases, in line with previous studies.[14,15] The effect of corrugation is evident. The inclusion of van der Waals interactions and exact HF exchange alters the energy difference between different phases but does not change the phase order.

**Table 1. Relative Stabilities of Beta $C_3N_4$ ($\beta$-$C_3N_4$) and Triazine- and Heptazine-Based g-$C_3N_4$ by Different Methods, after Geometry Optimization with (Wc) or without (Nc) Crystal Symmetry Constraints**

| $\triangle E$ (eV/$C_3N_4$) | PBE | HSE06 | PBE+D3 | HSE06+D3 |
|---|---|---|---|---|
| $\beta$-$C_3N_4$ | 2.100 | 1.502 | 1.706 | 1.180 |
| tri-$C_3N_4$-wc (flat) | 0.614 | 0.542 | 0.493 | 0.427 |
| tri-$C_3N_4$-nc | 0.392 | 0.372 | 0.316 | 0.284 |
| hep-$C_3N_4$-wc (flat) | 0.295 | 0.288 | 0.271 | 0.268 |
| hep-$C_3N_4$-nc | 0 | 0 | 0 | 0 |

**3.1.1. Beta Carbon Nitride.** As discussed in the computational details and shown in Table 1, we have performed full structural optimizations of $\beta$-$C_3N_4$ using the PBE, HSE06, PBE+D3, and HSE06+D3 DFT methods, respectively. The optimized lattice parameters, volumes, and band gaps are summarized in Table 2, together with previously reported experimental and DFT results for comparison. The experimental structural parameters (a = b = 6.42 Å, c = 2.43 Å, V = 86.74 Å$^3$)[48] provide a benchmark for evaluating the accuracy of the computational methods.

**Table 2. Lattice Parameters, Unit-Cell Volume, and Band Gap of Fully Optimized $\beta$-$C_3N_4$ (14 Atoms) Calculated Using Different Methods[c]**

| Bulk $\beta$-$C_3N_4$ | a | b | c | $\alpha$ | $\beta$ | $\gamma$ | Volume (Å$^3$) | Band gap[b] (eV) |
|---|---|---|---|---|---|---|---|---|
| EXP[48] | 6.42 | 6.42 | 2.43 | 90 | 90 | 120 | 86.74 | NA |
| LDA[14] | 6.40 | 6.40 | 2.40 | 90 | 90 | 120 | 85.13 | 3.3 |
| LDA($G_0W_0$)[a,20] | 6.41 | 6.41 | 2.41 | 90 | 90 | 120 | 85.76 | 4.9 |
| PBE | 6.44 | 6.44 | 2.44 | 90 | 90 | 120 | 87.56 | 3.2 |
| HSE06 | 6.38 | 6.38 | 2.41 | 90 | 90 | 120 | 84.98 | 4.9 |
| PBE+D3 | 6.42 | 6.42 | 2.43 | 90 | 90 | 120 | 86.64 | 3.2 |
| HSE06+D3 | 6.35 | 6.35 | 2.40 | 90 | 90 | 120 | 83.99 | 4.9 |

[a]NA denotes Not Available. [b]Band gap values by DFT methods are computed at their corresponding optimized geometries. [c]The LDA method for lattice parameters and $G_0W_0$ for band gap calculations.

Among all the methods used in this work, the PBE+D3 method predicts lattice parameters and volume that agree almost exactly with the experimental values.[48] In contrast, PBE slightly overestimates these structural parameters (a = b = 6.44 Å, c = 2.44 Å, V = 87.56 Å$^3$), a well-known behavior due to the lack of dispersion interactions, whereas HSE06 produces a more contracted structure with a volume of 84.98 Å$^3$. Although the inclusion of dispersion correction further increased the deviation in HSE06 (V = 83.99 Å$^3$), the results remained in close agreement with the experimental values.[48] Overall, all the methods used in this work reproduce the structural properties reasonably well relative to previous experimental[48] and DFT calculations.[14,20]

Figure 2 presents the calculated density of states (DOS) profiles of $\beta$-$C_3N_4$ obtained with the various methods considered in this work. These results enable a direct comparison of how different exchange−correlation functionals and the inclusion of dispersion corrections affect the electronic structure, especially the distribution of states near the valence- and conduction-band edges. Across all computational methods, the valence band maximum (VBM) is predominantly contributed by N atomic orbitals, whereas the conduction band minimum (CBM) retains strong C atomic orbital character, which is highly hybridized with N atomic orbitals. This consistent orbital distribution indicates that the bonding framework of $\beta$-$C_3N_4$ is robust against computational methods.

While the DOS profiles calculated by different methods are qualitatively similar, significant quantitative differences exist in the computed band gaps. Unfortunately, since it is a metastable bulk phase, it was not possible to find a reliable experimental band gap value. For this reason, we will consider the $G_0W_0$ value as our reference to assess the accuracy of the DFT methods.

The PBE functional predicts the smallest band gap (3.2 eV, see Table 2), reflecting its well-known tendency to underestimate the band gap due to the self-interaction error. The inclusion of D3 dispersion (PBE+D3) leads to a small structural modification that reflects in a negligible change in the DOS, with a band gap of 3.2 eV. In contrast, the HSE06 hybrid functional significantly enlarges the band gap to 4.9 eV. This widening is attributed to the introduction of nonlocal electron (exact) exchange, which effectively reduces the self-interaction and delocalization errors inherent in semilocal functionals.[49,50] The HSE06+D3 results further confirm this trend, showing a DOS profile nearly identical to that of HSE06, thereby reaffirming that dispersion corrections slightly affect the atomic geometry, with a resulting negligible effect on the electronic structure.

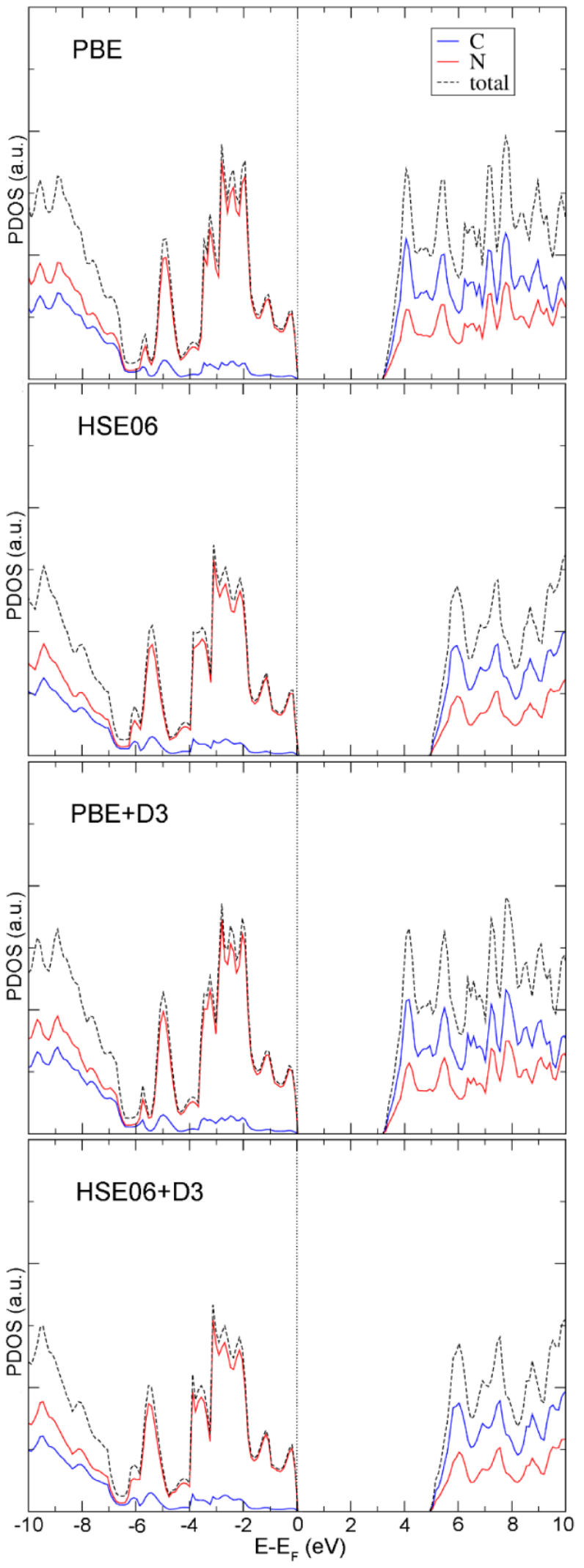

**Figure 2.** DOS of fully optimized $\beta$-$C_3N_4$ by different methods.

When benchmarked against the 4.9 eV band gap predicted by the more advanced $G_0W_0$ quasiparticle approach[20] (note that this value may not be a perfect reference since the starting calculation for the $G_0W_0$ was performed with the LDA functional), the HSE06-based calculations show the closest agreement, while both PBE and PBE+D3 significantly underestimate the band gap, indicating that hybrid functionals are essential for describing the electronic property of β-$C_3N_4$ with reasonable computational cost. In general, above comparative analysis reveals that while all four approaches predict similar orbital features in the DOS, only hybrid-functional-based methods successfully capture the wide-band gap nature of β-$C_3N_4$.

Table 3. Lattice Parameters, Unit-Cell Volume, and Band Gap of Triazine-Based g-$C_3N_4$ with/without Crystal Symmetry Constraints (21 Atoms) by Different Methods[c]

| Bulk triazine-based g-$C_3N_4$ | a | b | c | α | β | γ | Interlayer spacing (Å) | Volume (Å$^3$) | Band gap[b] (eV) |
|---|---|---|---|---|---|---|---|---|---|
| EXP[51] | NA | NA | NA | NA | NA | NA | NA | NA | 3.1 |
| LDA[14] | 4.74 | 4.74 | 10.08 | 90 | 90 | 120 | 3.36 | 196.13 | |
| PBE[54] | 4.79 | 4.79 | 10.16 | 90 | 90 | 120 | 3.39 | 201.88 | |
| PBE+TS[53] | 4.74 | 4.74 | 10.08 | 90 | 90 | 120 | 3.36 | 196.13 | |
| LDA($G_0W_0$)[a],[20] | 4.75 | 4.75 | 9.88 | 90 | 90 | 120 | 3.29 | 193.05 | 3.0 |
| with crystal symmetry constraints (tri-$C_3N_4$-wc) | | | | | | | | | |
| PBE | 4.78 | 4.78 | 10.79 | 90 | 90 | 120 | 3.60 | 213.87 | 1.5 |
| HSE06 | 4.74 | 4.74 | 10.53 | 90 | 90 | 120 | 3.51 | 205.01 | 3.0 |
| PBE+D3 | 4.78 | 4.78 | 9.44 | 90 | 90 | 120 | 3.15 | 186.55 | 1.3 |
| HSE06+D3 | 4.74 | 4.74 | 9.44 | 90 | 90 | 120 | 3.15 | 183.42 | 2.8 |
| no crystal symmetry constraints (tri-$C_3N_4$-nc) | | | | | | | | | |
| PBE | 4.71 | 4.71 | 11.37 | 80.37 | 92.51 | 120.01 | 2.92/2.87/2.90 | 215.32 | 2.3 |
| HSE06 | 4.69 | 4.68 | 10.87 | 91.67 | 90.40 | 120.18 | 2.80/2.94/2.84 | 206.22 | 3.7 |
| PBE+D3 | 4.71 | 4.71 | 9.98 | 89.65 | 90.67 | 120 | 2.53/2.45/2.53 | 191.92 | 2.2 |
| HSE06+D3 | 4.67 | 4.70 | 9.84 | 88.00 | 86.73 | 120.09 | 2.54/2.45/2.58 | 185.89 | 3.5 |

[a]The LDA method for lattice parameters and $G_0W_0$ for band gap calculations. [b]Band gap values by DFT methods are computed at their corresponding optimized geometries. [c]Regarding the interlayer spacing of tri-C3N4-nc, due to the corrugated layered structure, we have listed the minimum interlayer spacing between each layer.

**3.1.2. Graphitic Carbon Nitride Based on Triazine Units.** In Table 3, we present a comparative summary of the lattice parameters, volumes, and band gaps of triazine-based g-$C_3N_4$ obtained with the PBE, HSE06, PBE+D3, and HSE06+D3 methods, alongside previously reported experimental and DFT data.[14,20,51] The synthesized triazine-based g-$C_3N_4$ powders are amorphous, evidenced by a very broad XRD reflection centered at approximately 3.0 Å.[51] Due to the lack of well-defined experimental lattice parameters, we will adopt the PBE+TS (PBE augmented with dispersion corrections of Tkatchenko and Scheffler[52]) results[53] as the benchmark since the similar PBE+D3 method has been shown to provide a highly reasonable description of the lattice constants and volume relative to experimental data for β-$C_3N_4$,; see Table 2.

We must notice that in previous computational studies, bulk structures with flat layers have been mostly reported[14,15,20−26] (see Table S1 in the SI for more details). On the contrary, in the present work, we compare flat layers in crystal symmetry with models where no crystal symmetry constraints are imposed, allowing for the layers' corrugation or buckling. Indeed, we have found that the structure can further relax with some energy gain (see Table 1 for energetics and Table 2 for lattice parameters and band gap values).

Compared with the PBE+TS parameters (a = b = 4.74 Å, c = 10.08 Å, V = 196.13 Å$^3$),[53] the PBE functional severely overestimates the interlayer spacing (c = 11.37 Å) and predicts an expansion of the equilibrium volume to 215.32 Å$^3$ for the corrugated tri-$C_3N_4$-nc model. HSE06 reduces this overestimation, with a contracted c parameter (10.87 Å) and a modestly lower volume of 206.22 Å$^3$. Inclusion of D3 dispersion corrections significantly reduces the c lattice parameter for both semilocal and hybrid functionals. The PBE+D3 method effectively corrects the interlayer spacing (c = 9.98 Å) and predicts an equilibrium volume (191.92 Å$^3$) similar to PBE+TS, as one would expect. The structure predicted by HSE06+D3 is slightly more compact (V = 185.89 Å$^3$). Overall, the comparison reveals that dispersion interactions play a crucial role in describing the layered triazine-based g-$C_3N_4$.

Compared to the DOS characteristics of β-$C_3N_4$ discussed above (Figure 2), the DOS of triazine-based g-$C_3N_4$ (Figure 3) exhibits significant similarities but also some distinctions. In both cases, all four different computational methods obtain broadly similar orbital distributions, with N atomic orbitals dominating the VBM, while C atomic orbitals significantly contribute to the CBM and exhibit strong hybridization with N atomic orbitals. This similar orbital distribution across two independent structures highlights the robustness of the C−N covalent framework and indicates that the qualitative bonding characteristics of $C_3N_4$ are largely independent of the structural models and computational methods.

However, the differences among the methods are less pronounced when examining the DOS features and the positions of the band edges. In contrast to the DOS profiles of β-$C_3N_4$ (Figure 2), for which HSE06 predicts a substantial band gap enlargement relative to PBE, the DOS profiles of both the corrugated and flat triazine-based g-$C_3N_4$ show a smaller band gap contraction derived from PBE. The structural modification following the addition of D3 dispersion to PBE again causes only marginal adjustments in DOS peak positions and relative intensities, reinforcing that dispersion corrections have limited impact. In comparison, the HSE06 and HSE06+D3 methods slightly modify the shape of the DOS peaks and produce quite more separated band edges.

When comparing the band gap of corrugated triazine-based g-$C_3N_4$ (tri-$C_3N_4$-nc) with the experimentally reported value of 3.1 eV, the deviations among the computational methods become less pronounced than those of the β-$C_3N_4$ phase analyzed above. Both PBE and PBE+D3 calculations of triazine-based g-$C_3N_4$ generate moderate underestimations, with band gaps of 2.3 and 2.2 eV, respectively. In contrast, the HSE06 functional overestimates the band gap (3.7 eV) compared to the experimental value (3.1 eV), although the inclusion of dispersion interactions in the geometry optimization alleviates this case (3.5 eV for HSE06+D3). As summarized in Table 3, the $G_0W_0$ method[20] predicted the band gap value (3.0 eV) closest to the experiment, establishing it as the most accurate benchmark. However, HSE06+D3 achieves similar results with significantly lower computational costs than $G_0W_0$, making it an excellent balanced approach for this system.

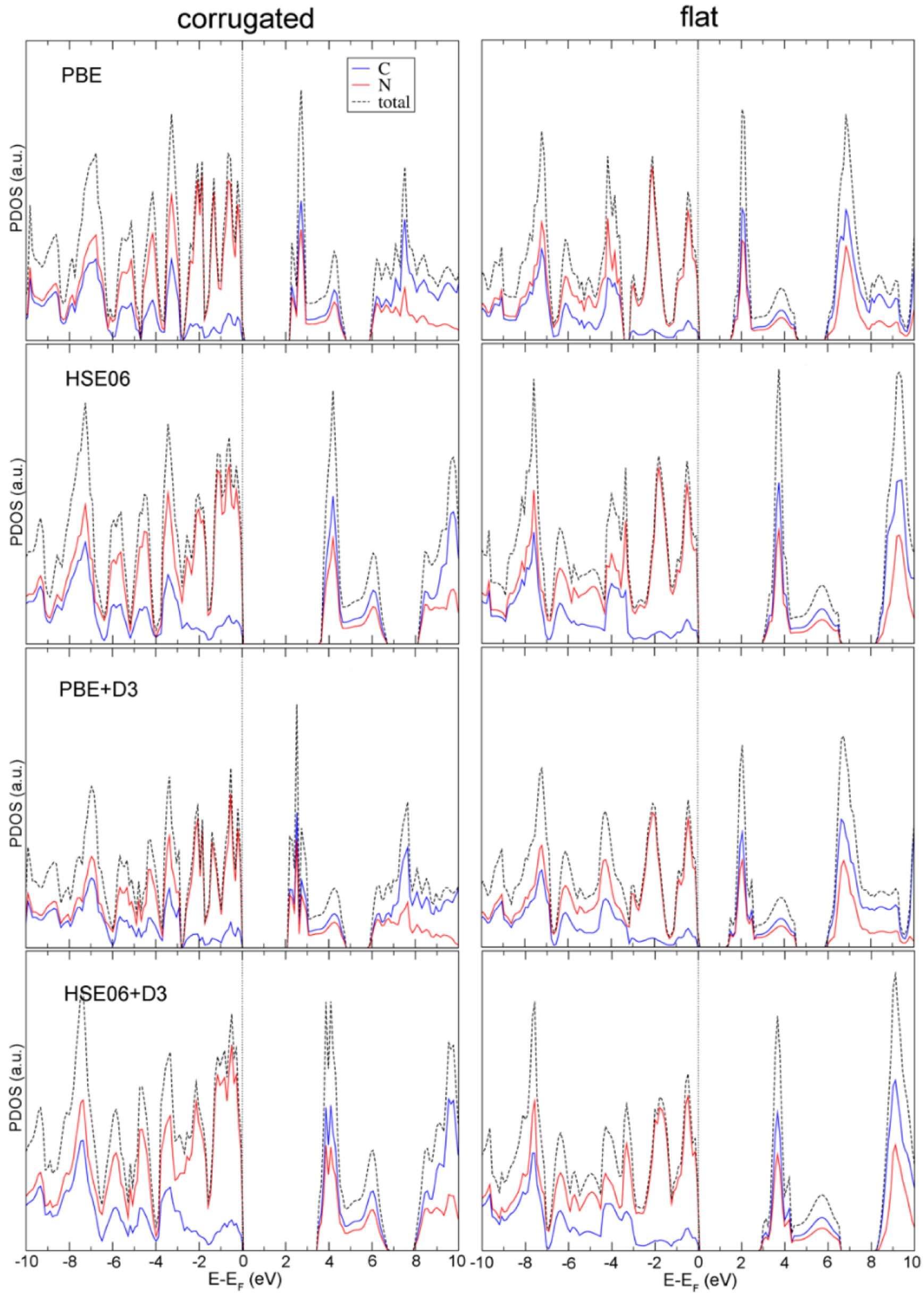

**Figure 3.** DOS of corrugated (tri-$C_3N_4$-nc) and flat (tri-$C_3N_4$-wc) triazine-based g-$C_3N_4$ by different methods.

Comparing the DOS of the corrugated tri-$C_3N_4$-nc structure with that of the flat tri-$C_3N_4$-wc one illustrated in Figure 3, we can observe some pronounced differences. The flat configuration is characterized by a significantly larger band gap than that of its corrugated counterpart, and this trend is essentially independent of the computational method employed, with the detailed values summarized in Table 3. For the flat tri-$C_3N_4$-wc structure, the HSE06 functional predicts a band gap of about 3.0 eV, which is in very good agreement with the experimental value of 3.1 eV[51] and the $G_0W_0$ calculated band gap.[20] In contrast, the HSE06 functional predicts a larger band gap (3.7 eV) for the corrugated tri-$C_3N_4$-nc structure shown in Figure 3, indicating a slight overestimation of the band gap. Including D3 dispersion into HSE06 slightly alleviates this case, reducing the band gaps to 2.8 eV for tri-$C_3N_4$-wc and 3.5 eV for tri-$C_3N_4$-nc model.

**3.1.3. Graphitic Carbon Nitride Based on Heptazine Units.** Table 4 shows a systematic comparison of the lattice parameters, equilibrium volumes, and band gaps of heptazine-based g-$C_3N_4$ obtained using different exchange−correlation functionals, in conjunction with experimental and DFT results from the literature.[2,20,26,55] Experimental studies on corrugated polymeric phases report in-plane lattice constants of a = b ≈ 6.8−7.0 Å and an out-of-plane lattice constant c = 6.5 Å, characteristic of a compact framework.[2,55] In contrast, PBE predicts a corrugated structure for hep-$C_3N_4$-nc and significantly overestimates out-of-plane lattice constant c (a = b = 6.94 Å, c = 7.76 Å), leading to an equilibrium volume expansion to 313.57 Å$^3$, consistent with the known underbinding tendency of semilocal functionals, especially in layered materials. Introducing dispersion corrections (PBE+D3) brings the optimized structure closer to the experiment (Table 4). The PBE+D3 geometry (a = b = 6.95 Å, c = 7.10 Å, V = 284.28 Å$^3$) greatly reduces the interlayer distance, resulting in a much more compact structure relative to PBE. The HSE06 functional further shrinks the in-plane lattice (a = b = 6.88 Å) but simultaneously expands the c parameter (8.19 Å), leading to a moderately reduced volume of 303 Å$^3$. When dispersion is included (HSE06+D3), both the in-plane and out-of-plane lattice constants become close to experimental values (a = b = 6.90 Å, c = 6.99 Å, V = 280.33 Å$^3$), representing the closest agreement among all the methods used in this work. For comparison, the structural parameters predicted by previous PBE+D3 calculations (a = b = 6.99 Å, c = 6.43 Å, V = 265.47 Å$^3$)[26] are close to the experimental values.[2,55] Meanwhile, previously reported $G_0W_0$ calculations based on a different structural model generate markedly anisotropic lattice parameters and a much larger volume (a = 7.08 Å, b = 12.27 Å, c = 6.87 Å, V = 596.81 Å$^3$), highlighting the structural sensitivity of g-$C_3N_4$ to the choice of model. Overall, our calculations demonstrate that both nonlocal electron exchange and dispersion corrections are essential for capturing the experimentally observed structure of heptazine-based g-$C_3N_4$.

**Table 4. Lattice Parameter, Volume, and Band Gap of Heptazine-Based g-$C_3N_4$ with/without Crystal Symmetry Constraints (28 Atoms) by Different Methods[c]**

| Bulk heptazine-based g-$C_3N_4$ | a | b | c | $\alpha$ | $\beta$ | $\gamma$ | Interlayer spacing (Å) | Volume (Å$^3$) | Band gap[b] (eV) |
|---|---|---|---|---|---|---|---|---|---|
| EXP-corrugated polymer[2] | 6.81 | 6.81 | 6.5 | | | | 3.25 | | 2.7 |
| EXP-corrugated polymer[55] | 6.97 | 6.97 | 6.5 | | | | 3.25 | | 2.7 |
| PBE+D3[26] | 6.99 | 6.99 | 6.43 | 77.58 | 96.17 | 120.01 | NA | 265.47 | 1.7 |
| LDA($G_0W_0$)-different structure[a,20] | 7.08 | 12.27 | 6.87 | 90 | 90 | 90 | 3.44 | 596.81 | 2.9 |
| with crystal symmetry constraints (hep-$C_3N_4$-wc) | | | | | | | | | |
| PBE | 7.13 | 7.13 | 7.12 | 90 | 90 | 120 | 3.56 | 313.82 | 1.1 |
| HSE06 | 7.07 | 7.07 | 7.04 | 90 | 90 | 120 | 3.52 | 304.53 | 2.6 |
| PBE+D3 | 7.12 | 7.12 | 6.44 | 90 | 90 | 120 | 3.22 | 283.04 | 1.0 |
| HSE06+D3 | 7.06 | 7.06 | 6.46 | 90 | 90 | 120 | 3.23 | 279.02 | 2.2 |
| no crystal symmetry constraints (hep-$C_3N_4$-nc) | | | | | | | | | |
| PBE | 6.94 | 6.94 | 7.76 | 98.22 | 96.39 | 119.69 | 2.35/2.36 | 313.57 | 1.9 |
| HSE06 | 6.88 | 6.88 | 8.19 | 106.84 | 98.21 | 119.72 | 2.31/2.31 | 303.00 | 3.0 |
| PBE+D3 | 6.95 | 6.95 | 7.10 | 105.02 | 90 | 119.78 | 2.08/2.10 | 284.28 | 1.8 |
| HSE06+D3 | 6.90 | 6.90 | 6.99 | 83.33 | 82.71 | 119.79 | 2.11/2.11 | 280.33 | 2.9 |

[a]The LDA method for lattice parameters and $G_0W_0$ for band gap calculations. [b]Band gap values by DFT methods are computed at their corresponding optimized geometries. [c]Regarding the interlayer spacing of hep-C3N4-nc, due to the corrugated layered structure, we have listed the minimum interlayer spacing between each layer.

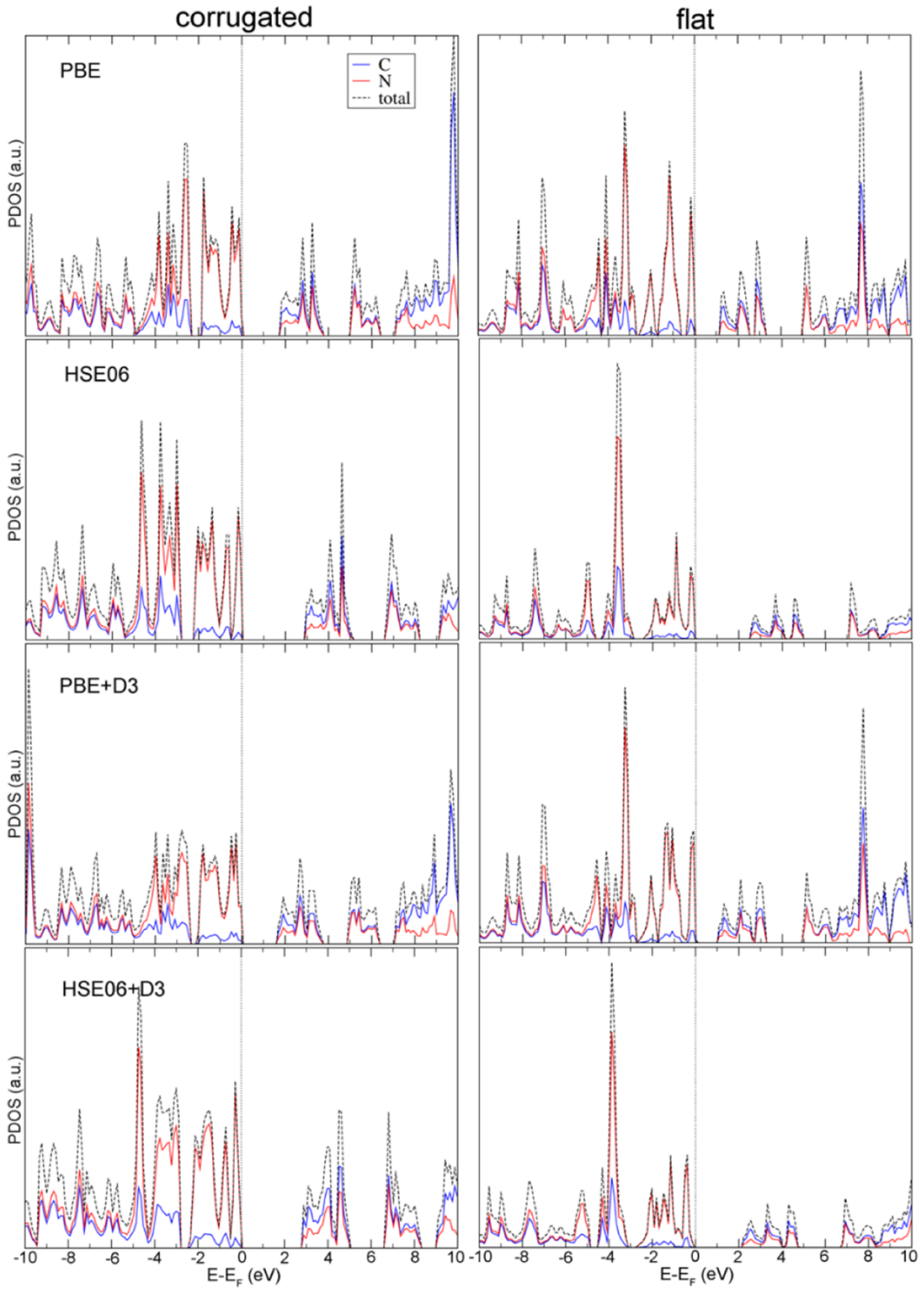

**Figure 4.** DOS of corrugated (hep-$C_3N_4$-nc) and flat (hep-$C_3N_4$-wc) heptazine-based g-$C_3N_4$ by different methods.

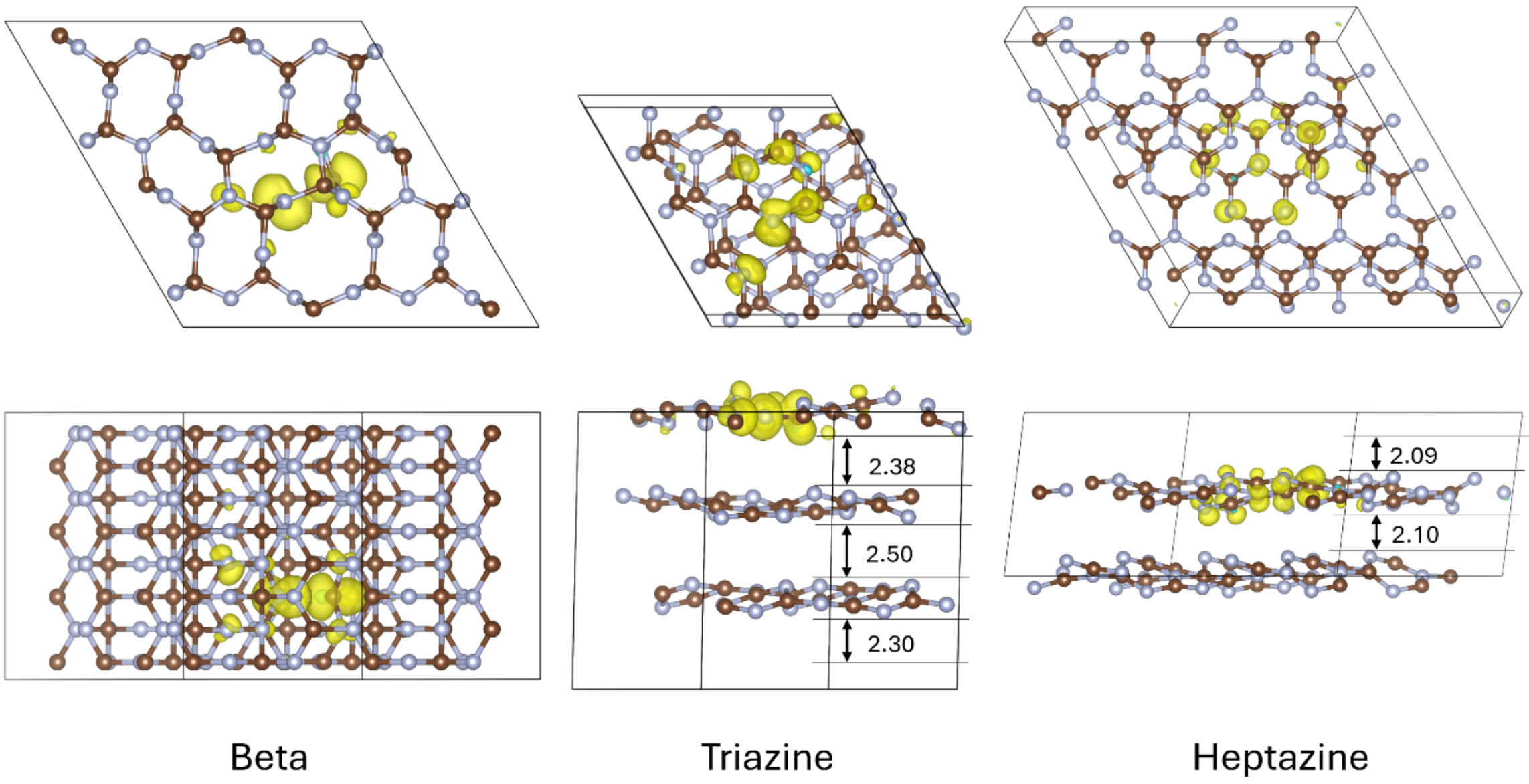

**Figure 5.** Spin plots of the optimized triplet exciton in bulk $C_3N_4$ calculated by the HSE06+D3 method. The isovalue used for the contour plots is 0.01 e/Å$^3$. The brown and gray spheres represent carbon and nitrogen atoms, respectively.

Compared to the DOS features of $\beta$-$C_3N_4$ (Figure 2) and triazine-based g-$C_3N_4$ (Figure 3), the DOS profiles of heptazine-based g-$C_3N_4$ (Figure 4) show broadly similar orbital contributions but exhibit stronger peak enrichment and more pronounced hybridization characteristics due to its larger $\pi$-conjugated heptazine units. As in the previous two structures, all four computational methods consistently indicate that N atomic orbitals dominate the upper valence-band region, whereas C atomic orbitals contribute primarily to the conduction-band states, maintaining strong C−N hybridization. This consistent electronic distribution across the three different $C_3N_4$ structures clearly demonstrates the robustness of the C−N bonding framework and indicates that the qualitative orbital character of $C_3N_4$ is largely independent of the structural model and the computational methods.

In contrast with the $\beta$-$C_3N_4$ (Figure 2) and triazine-based g-$C_3N_4$ (Figure 3), the heptazine-based g-$C_3N_4$ phase (Figure 4) exhibits the weakest functional-dependent bandgap variation. The band gap difference predicted by PBE and HSE06 decreases from 1.8 eV for $\beta$-$C_3N_4$ to 1.4 eV for corrugated triazine-based g-$C_3N_4$ (tri-$C_3N_4$-nc) and further to only 1.1 eV for corrugated heptazine-based g-$C_3N_4$ (hep-$C_3N_4$-nc) (see Tables 2, 3, and 4 for the detailed values). This reduced sensitivity indicates that the electronic structure of corrugated heptazine-based g-$C_3N_4$ is less affected by self-interaction errors. Besides, the inclusion of D3 dispersion corrections introduces tiny changes in the structural features that reflect in very similar DOS profiles.

When comparing the band gap of corrugated heptazine-based g-$C_3N_4$ (hep-$C_3N_4$-nc) with the experimentally reported value of 2.7 eV[2,55] Figure 4 indicates that PBE and PBE+D3 calculations significantly underestimate the band gap, obtaining values of 1.9 and 1.8 eV, respectively. In contrast, hybrid functional calculations provide systematic improvement. HSE06 predicts a band gap of 3.0 eV, while HSE06+D3 slightly lowers it to 2.9 eV, in excellent agreement with the $G_0W_0$ result[20] and closely matching the experimental measurement.[2,55] Overall, the results indicate that HSE06+D3 achieves the best balance between accuracy and computational cost in describing the band gap.

Similar to the DOS difference between the corrugated and flat models of triazine-based g-$C_3N_4$ (Figure 3), the DOS of heptazine-based g-$C_3N_4$ shown in Figure 4 also clearly indicates that the flat configuration (hep-$C_3N_4$-wc) exhibits a significantly larger band gap than its corrugated counterpart (hep-$C_3N_4$-nc). This trend remains consistent regardless of the computational method used, with the detailed values provided in Table 4. HSE06 predicts band gaps of 2.6 and 3.0 eV for hep-$C_3N_4$-wc and hep-$C_3N_4$-nc, respectively, both close to the experimental value of 2.7 eV[2,55] and the $G_0W_0$ predicted value of 2.9 eV.[20] When D3 corrections are included, the resulting band gap of hep-$C_3N_4$-wc is underestimated to a value of 2.2 eV. By contrast, as the most stable configuration, hep-$C_3N_4$-nc obtains a band gap of 2.9 eV with HSE06+D3, again in close agreement with the experimental values.[2,55] In summary, the corrugated structures of heptazine-based g-$C_3N_4$ are the most stable configurations and, as evaluated by the HSE06+D3 method, exhibit band gaps that closely align with the experimental values.

**3.1.4. Photoexcited Bulk Phase.** As mentioned in the introduction, carbon nitrides are promising photocatalytic materials that undergo photoexcitation upon vis-light irradiation.[16,56−58] The formation of spin triplet states plays a crucial role in the photochemistry of a material because of their relatively long lifetime, which favors their involvement in surface chemical reactions.[59−63] Triplet excitons can more easily evolve into free charge carriers that drive redox reactions. Therefore, we consider it particularly relevant to investigate, through spin-constrained calculations, the nature of triplet excitons in the different $C_3N_4$ phases, with special focus on the extent of their localization/delocalization through the crystal lattice.[43] This can be achieved through a spin-constrained optimization in the triplet state, resulting in an electron hole trapped at the top of the valence band and a photoexcited electron trapped at the bottom of the conduction band.[64] For this specific calculation, we have used a bulk supercell model ($2 \times 2 \times 4$ for $\beta$-$C_3N_4$ and $2 \times 2 \times 1$ for the triazine-based and $2 \times 2 \times 1$ for the heptazine-based g-$C_3N_4$) in order to have the proper description of the spin localization/delocalization. In Figure 5, we present the spin density plot of the fully optimized self-trapped triplet excitons in the three bulk structures considered in this work. We wish to note that the spin density plot is the sum of the spatial distribution of the two unpaired electrons of the triplet state: one is the photoexcited electron, and the other is the electron that is left unpaired by the presence of the photoexcited hole in the same electronic state. Therefore, the spatial distribution of this second unpaired electron is a reliable representation of the spatial distribution of the corresponding hole. For this reason, we can consider the spin density plot as the spatial representation of the triplet exciton (electron+hole).

The spin density plot (yellow) for the optimized triplet exciton in $\beta$-$C_3N_4$ is characterized by a strong localization on a pair of specific C (spin = 0.74) and N (spin = 0.82) atoms, whose chemical bond is found to be broken, with a distance of 2.04 Å compared to 1.44 Å for C−N in the $\beta$-$C_3N_4$ ground state structure. In Table 5, we report the energy difference per cell between the $S_0$ ground state and $T_1$ for bulk $\beta$-$C_3N_4$ (+5.01 eV) and the relaxation energy of the $T_1$ spin configuration, leading to the triplet exciton self-trapping through the bulk lattice relaxation (−0.59).

**Table 5. Total Energy Variation (HSE06+D3) between $S_0$ and $T_1$ (Band Gap Value from Orbital Energy Difference in Parentheses for Comparison) at the Ground State Geometry, Full Atomic Relaxation Energy of $T_1$ Configuration (i.e., Exciton Trapping Energy), and Vertical Energy Variation between Fully Relaxed $T_1$ and $S_0$**

| $\Delta E$ (eV/cell) | $S_0 \rightarrow T_1$ | $T_1 \rightarrow T_{1_OPT}$ | $T_{1_OPT} \rightarrow S_0$ |
|---|---|---|---|
| $\beta$-$C_3N_4$ | 5.01 (4.9) | −0.59 | −1.76 |
| tri-$C_3N_4$-nc | 3.61[a] (3.5)[b] | −0.53 | −2.27 |
| hep-$C_3N_4$-nc | 2.87 (2.9) | −0.19 | −2.51 |

[a]This ΔSCF energy difference was computed with a larger $4 \times 4 \times 1$ supercell model. [b]This orbital energy difference is the same for $4 \times 4 \times 1$ and $2 \times 2 \times 1$ supercell models.

In the case of triazine-based bulk g-$C_3N_4$, the energy difference (ΔSCF) from the ground state $S_0$ to the excited state $T_1$ (3.61 eV) is very close to the band gap value of this system, as obtained from the orbital energy difference in the PDOS calculations (3.5 eV), although slightly larger. This unexpected discrepancy is probably due to the fact that one is a total energy difference, whereas the second is the difference between Kohn−Sham eigenvalues. After atomic relaxation, there is an energy gain of −0.53 eV due to some polaronic effects. The triplet exciton is found to be mostly delocalized on the atoms of a single triazine unit.

Similarly, in the case of heptazine-based bulk g-$C_3N_4$, the energy difference (ΔSCF) from the ground state $S_0$ to the excited state $T_1$ (2.87 eV) is very close to the band gap value as obtained from the orbital energy difference in the PDOS calculations (2.9 eV). After atomic relaxation, there is an energy gain of only −0.19 eV due to very tiny polaronic effects. The triplet exciton is found to be delocalized on the atoms of a single heptazine unit. To identify the partial contribution from the hole and the electron, we performed two additional calculations, where we either removed one electron or added one electron to the system and fully optimized to self-trap the extra positive ($h^+$) or negative ($e^-$) charge. The results are shown in Figure S1. We can observe that the hole is localized on the six edge N atoms of one heptazine unit, whereas the electron is more delocalized, with contributions from the p state of the central N atom of the heptazine unit and from the p states of the C atoms. When the hole and electron become paired in the self-trapped exciton, the electrostatic attraction (exciton binding) further localizes the electron on the same single heptazine unit as the hole.

From the analysis above, we may conclude that the three carbon nitride phases present quite different properties with respect to the exciton structural details and excitation/trapping energies. Bulk $\beta$-$C_3N_4$ is characterized by a larger gap and, therefore, by a larger excitation energy, followed also by a much larger exciton self-trapping energy that results from the breaking of one C−N bond and the formation of two radical species at 2.4 Å apart. The layered graphitic-like systems present a more similar behavior although the exciton localization/delocalization is strictly dependent on the repeating structural unit. This is because the exciton delocalization follows the $\pi$-conjugation of the layer and, therefore, is limited at the N-junctions between building block units. For this reason, the triplet exciton is found to delocalize in one single triazine unit for the triazine-based graphitic $C_3N_4$, whereas it is found to be delocalized in one single heptazine unit for the heptazine-based graphitic $C_3N_4$.

Finally, in order to prove the validity of our approach for the calculation of bulk triplet excitons, we have compared the computed energy difference for the $T_1 \rightarrow S_0$ transitions (through delayed fluorescence or phosphorescence mechanisms) with the experimental photoluminescence emission measurements reported in the literature[65−67] for g-$C_3N_4$ that are found to be in the range of 440−470 nm, thus corresponding to energies between 2.6 and 2.8 eV. The agreement with the values in Table 5 is very good: 2.27 eV for tri-$C_3N_4$-nc and 2.51 eV for hep-$C_3N_4$-nc; therefore, our model of triplet excitons based on HSE06-D3 calculation is robust.

## 3.2. Nanostructuring

**3.2.1. Nanoparticles of Beta Carbon Nitride.** The nanostructuring of a 3D bulk such as $\beta$-$C_3N_4$ can be obtained by preparing 0D nanoparticles. We have modeled a NP of 2 nm diameter. Surface atoms have been saturated with a corresponding number of H atoms to achieve 4-fold coordination for C and 3-fold coordination for N. The atomic and electronic structures of the hydrogen-saturated $C_3N_4$ nanoparticle are shown in Figure 6, and its bond lengths are listed in Table 6. Structurally, these nanoparticles retain the basic characteristics of the $\beta$-$C_3N_4$ framework, with an average C−N bond length of 1.44 Å. From the image, one can see that the edge C and N atoms are hydrogen-saturated through C−H (1.09 Å) and N−H (1.00 Å) bonds, effectively eliminating dangling bonds and stabilizing the system.

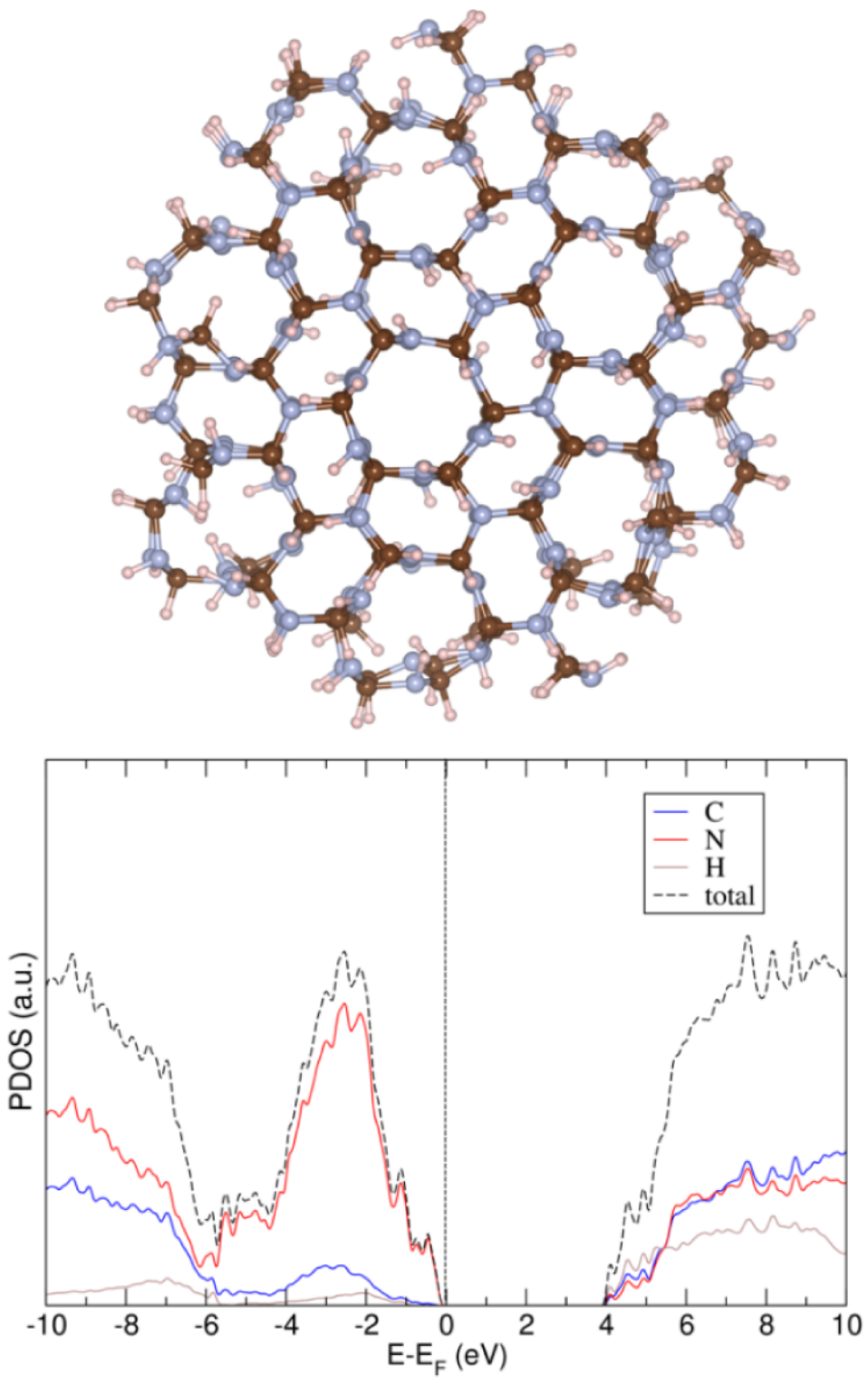

**Figure 6.** Atomic structure and DOS of hydrogen-saturated $C_3N_4$ nanoparticles calculated by the HSE06+D3 method. The brown, gray, and pink spheres represent carbon, nitrogen, and hydrogen atoms, respectively.

**Table 6. Bond Length of Hydrogen-Saturated $C_3N_4$ Nanoparticle Optimized by the HSE06+D3 Method**

| Average bond length (Å) | C−N | C−H | N−H |
|---|---|---|---|
| $C_3N_4$ nanoparticle | 1.44 | 1.09 | 1.00 |

Regarding electronic properties, the DOS results (Figure 6) show that the hydrogen-saturated $C_3N_4$ nanoparticle exhibits significant semiconductor characteristics, with a band gap of approximately 4.3 eV. Compared to the band gap of 4.9 eV for the corresponding three-dimensional bulk $\beta$-$C_3N_4$ (see Table 2 and Figure 2), the band gap of the nanoparticle system is significantly reduced. This band gap contraction primarily stems from the modulation of the electronic structure by edge hydrogen saturation. For the three-dimensional bulk $\beta$-$C_3N_4$, the VBM is mainly contributed by N atomic orbitals, while the CBM is contributed by both C and N atomic orbitals (see Figure 2). In the case of the nanoparticle, we observe an important modification of the CBM with the contribution from the H atoms (antibonding C−H and N−H states), which is the origin of the band gap reduction (Figure 6).

**3.2.2. Cleavage Energies of Triazine- and Heptazine-Based g-$C_3N_4$.** In this section, we investigate the cleavage energies of triazine- and heptazine-based g-$C_3N_4$ computed with different functionals and according to two different approaches.[68,69] Values in Table 7 indicate that the inclusion of vdW interactions plays a decisive role in predicting the exfoliation energy. Upon the introduction of the D3 dispersion corrections, the cleavage energies for both phases increase up to >20 meV/Å$^2$. Specifically, PBE+D3 and HSE06+D3 functionals predict that the cleavage energies of triazine-based g-$C_3N_4$ are 21.61 and 24.62 meV/Å$^2$, respectively, close to 20.66 meV/Å$^2$ by PBE+D3 and 23.10 meV/Å$^2$ by HSE06+D3 for heptazine-based g-$C_3N_4$, with the computational approach proposed in ref 68 (see Table 7). When we use the other approach proposed in ref 69, we obtain very similar values for HSE06+D3.

**Table 7. Cleavage Energies of Triazine- and Heptazine-Based g-$C_3N_4$ by Different Methods[c]**

| Cleavage energy (meV/Å$^2$) | PBE[a] | HSE06[a] | PBE+D3[a] | HSE06+D3[a] | HSE06+D3[b] |
|---|---|---|---|---|---|
| triazine-based g-$C_3N_4$ | 15.38 | 4.37 | 21.61 | 24.62 | 21.97 |
| heptazine-based g-$C_3N_4$ | 6.33 | 6.95 | 20.66 | 23.10 | 22.22 |

[a]Cleavage energy computed with the same method as in ref 68. [b]Cleavage energy computed with the same method as in ref 69. [c]All the structures are fully optimized without crystal symmetry constraints.

The above analysis clearly suggests that vdW interactions serve as the dominant driving force in maintaining the structural stability of the layered g-$C_3N_4$. Notably, the cleavage energy values of g-$C_3N_4$ are highly comparable to those of other two-dimensional materials, such as graphene and hexagonal boron nitride (h-BN).[68,70−72] These small cleavage energies indicate that the monolayer of g-$C_3N_4$ can be easily exfoliated, which agrees well with previous experimental results.[73−75]

As shown in Figure 7, the fully optimized triazine-based g-$C_3N_4$ structures (without symmetry constraints) exhibit an intrinsically corrugated layered geometry rather than an ideal planar configuration. The degree of corrugation is an inherent structural feature and is independent of the number of layers. When the layer number increases from monolayer to bilayer and trilayer, the stacking of adjacent layers introduces interlayer coupling and dispersion interactions, thereby enhancing the structural stability of the multilayer models rather than significantly altering the corrugated amplitude of each layer. This interlayer coupling energetically stabilizes the stacked configurations and favors the trilayer model, as proved by the relative stability calculations in Table 7.

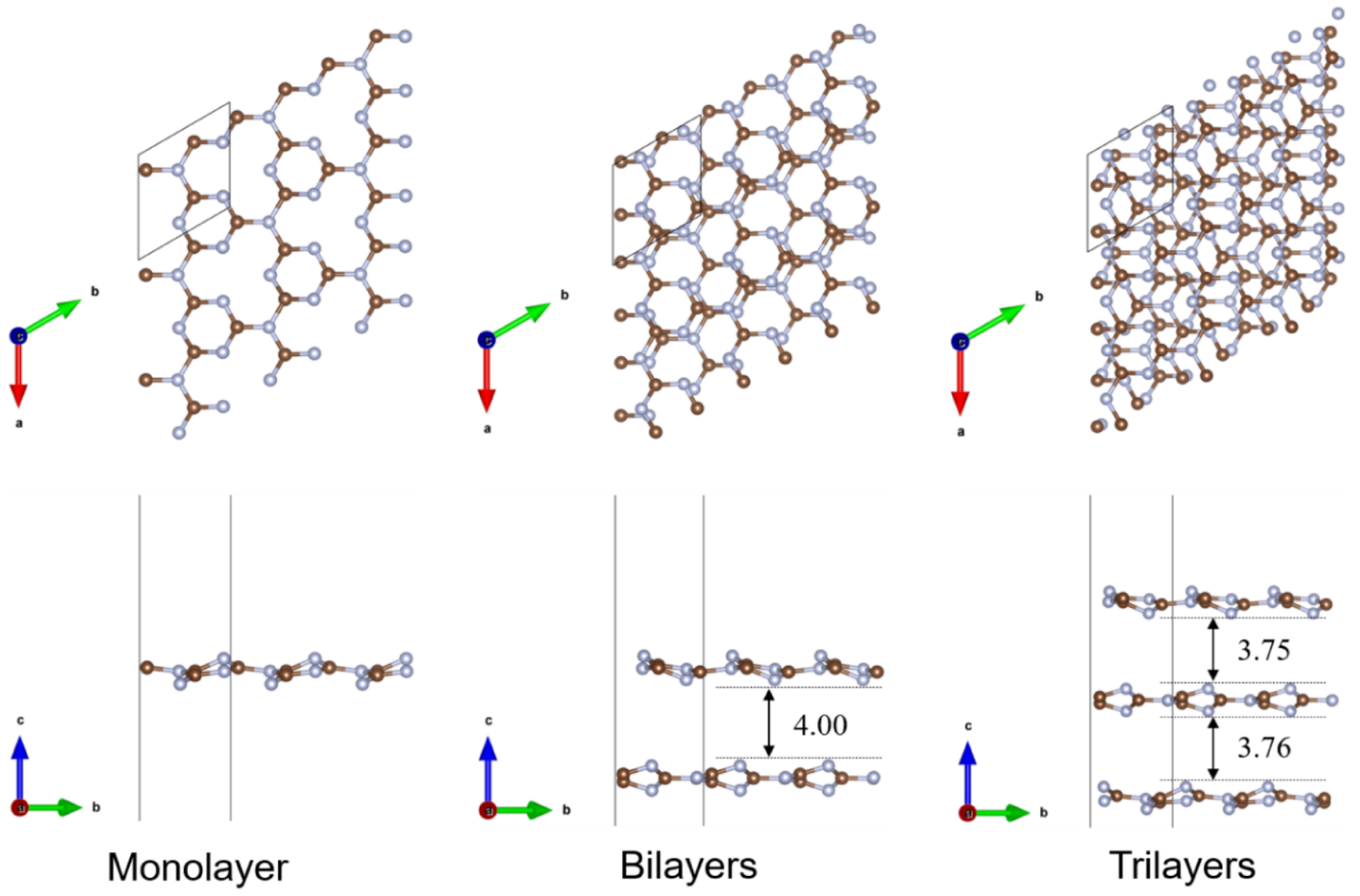

**Figure 7.** Atomic structure of 2D triazine-based g-$C_3N_4$ fully optimized without crystal symmetry constraints using the HSE06+D3 method.

The DOS spectra in Figure S2 clearly demonstrate that the electronic structure is significantly dependent on the layer thickness. The band gap progressively narrows with increasing layer number, from 3.9 eV for the monolayer to 3.7 eV for the bilayer and further to 3.6 eV for the trilayer, gradually approaching the bulk band gap value of 3.5 eV, as given in Table 3. This continuous band gap reduction is mainly attributed to enhanced interlayer interactions. Although the band gap decreases with increasing layer thickness, all layer configurations remain wide-band gap semiconductors, indicating that triazine-based g-$C_3N_4$ retains its semiconductor properties even in multilayer configurations.

Compared with the layered triazine-based g-$C_3N_4$, the heptazine-based g-$C_3N_4$ exhibits a similar trend in stability (Table 7). The bilayer model is energetically more favorable than the monolayer model across all computational methods. When dispersion correction is introduced, the energy difference between the monolayer and bilayer models further increases, which again confirms the significant role of interlayer dispersion interactions in terms of structural stability.

Figure 8 shows the atomic structure of heptazine-based g-$C_3N_4$ fully optimized using the HSE06+D3 method (without symmetry constraints). Similar to the triazine-based system, the heptazine-based g-$C_3N_4$ also exhibits a nonplanar, wrinkled layered configuration, with more pronounced corrugation, in line with previous calculations.[76] This enhanced corrugation is mainly attributed to its larger heptazine units and larger pores between these units. The increased size of the structural units and pores weakens the in-plane constraints, providing greater freedom for out-of-plane deformation, thus enhancing the overall corrugation of the two-dimensional network. Upon the formation of bilayer structures, the main stabilizing contribution of the system still comes from the interlayer dispersion interactions, and the resulting reduction in total energy is clearly reflected in Table 7.

Compared to the DOS of triazine-based g-$C_3N_4$ (Figure S2), heptazine-based g-$C_3N_4$ exhibits a similar but less distinct trend (Figure S3). Its band gap narrows slightly from 3.1 eV for the monolayer to 3.0 eV for the bilayer, gradually approaching the 2.9 eV of the bulk counterpart listed in Table 4. This relatively small band gap change indicates that the interlayer electronic coupling in the heptazine-based system is weaker than that in triazine-based g-$C_3N_4$.

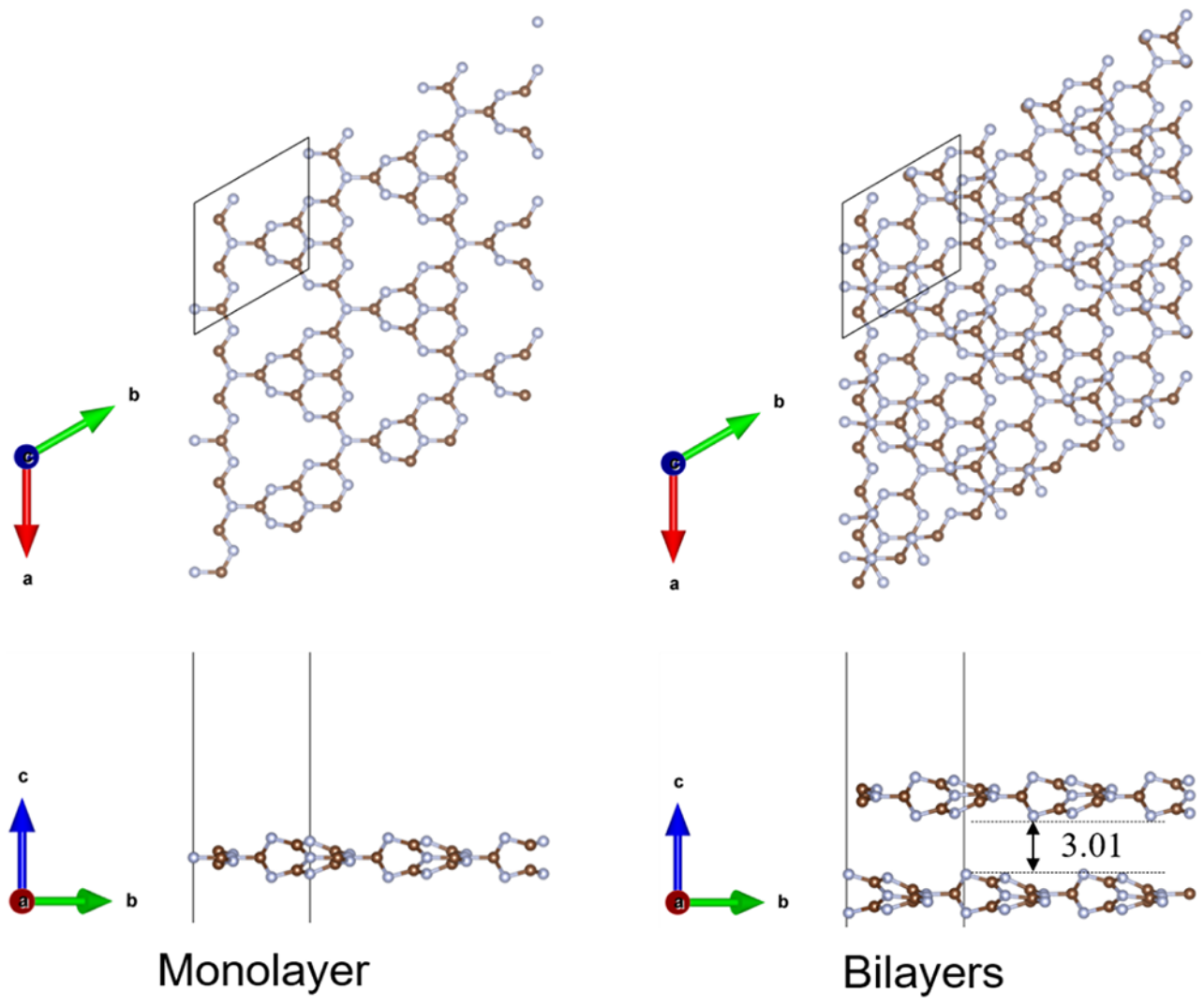

**Figure 8.** Atomic structure of fully optimized 2D heptazine-based g-$C_3N_4$ without crystal symmetry constraints using the HSE06+D3 method.

### 3.3. Doping of Nanostructures by Sulfur Atoms

Finally, in this section, we present the effect of S-doping when one S atom is introduced into the nanostructured models (0D and 2D). In the case of the $\beta$-$C_3N_4$ nanoparticles (0D), the substitution of one surface N(H) atom with one S atom leads to a sulfur concentration of about 0.4% by weight.

The C−S distances are 1.86 Å, which is quite larger than the C−N distances of 1.44 Å (Table 8), which is related to the larger atomic radius and different bonding characteristics of the S atom. In the DOS shown in Figure 9, we can observe a similarity with Figure 6 except for the projection in green on the sulfur atom. This projection has been magnified by ten times in order to highlight both the bonding (in the valence band) and antibonding (at the bottom of the conduction band) contributions of the S states. Moreover, S doping further slightly reduces the band gap to 4.2 eV compared to 4.3 eV for the undoped hydrogen-saturated $C_3N_4$ nanoparticles (Figure 6). Regarding the composition of the band-edge states, S atomic orbitals primarily contribute to the CBM, which undergoes a marked reconstruction, shifting from one primarily dominated by N atomic orbitals in the undoped case (Figure 6) to a combination of contributions from multiple atomic orbitals in the S-doped case (Figure 9). This change indicates that S doping alters the electronic properties of $\beta$-$C_3N_4$ nanoparticles by modulating the orbital hybridization for CBM.

**Table 8. Bond Length of Hydrogen-Saturated $C_3N_4$ Nanoparticle with Sulphur Doping Optimized by the HSE06+D3 Method**

| Average bond length (Å) | C−N | C−S | C−H | N−H |
|---|---|---|---|---|
| $C_3N_4$ nanoparticle | 1.44 | 1.86 | 1.09 | 1.00 |

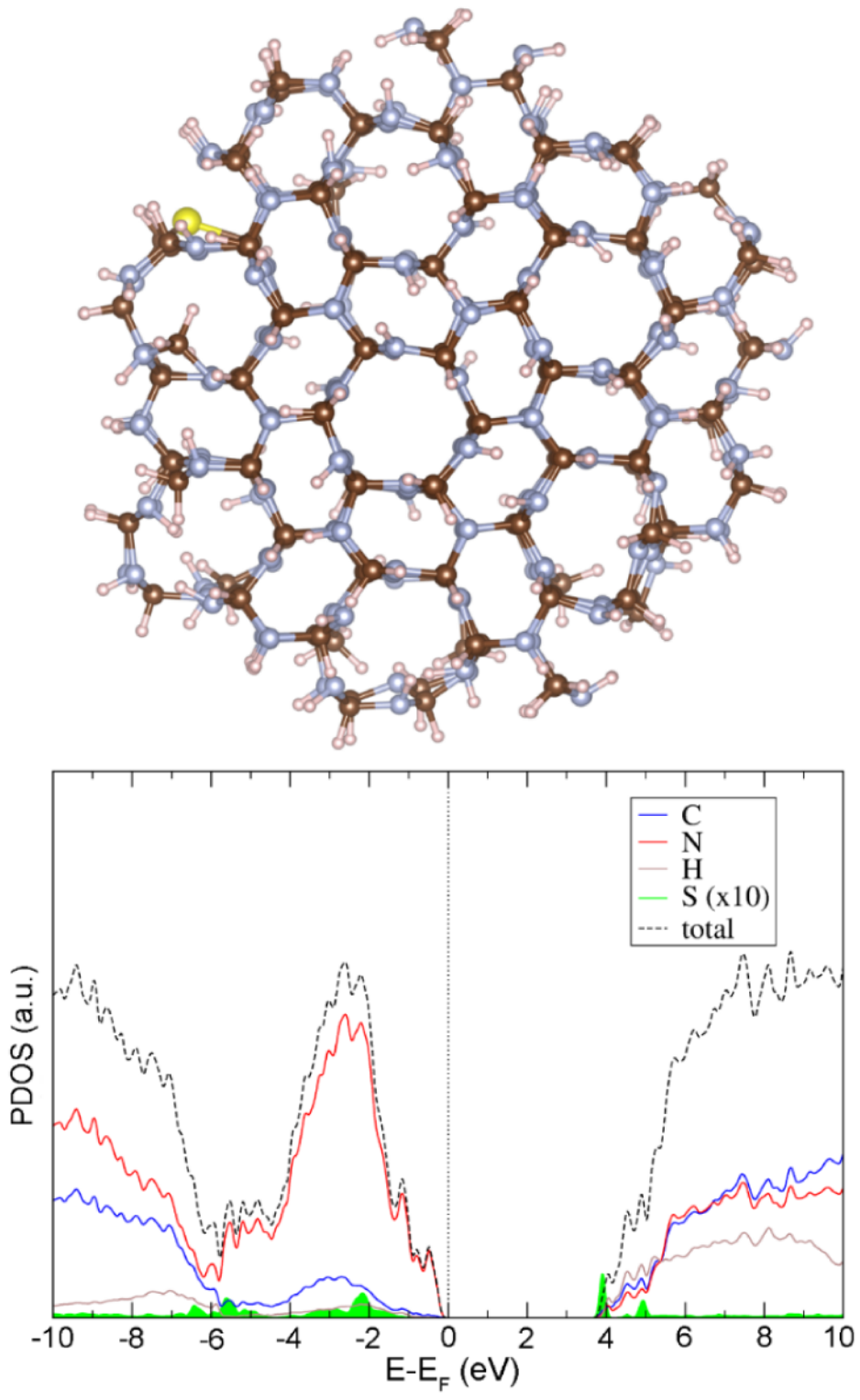

**Figure 9.** Atomic structure and DOS of hydrogen-saturated $C_3N_4$ nanoparticles with sulfur doping calculated by the HSE06+D3 method. The brown, gray, pink, and yellow spheres represent carbon, nitrogen, hydrogen, and sulfur atoms, respectively. The contribution of the S atomic orbitals to the DOS is multiplied by ten for clarity.

Overall, as illustrated in Figures 2, 6, and 9, the hydrogen-saturated $C_3N_4$ nanoparticle exhibits a significantly narrower band gap compared to the three-dimensional bulk $\beta$-$C_3N_4$ and has a different orbital composition for the CBM. Furthermore, S doping further reduces the band gap of the hydrogen-saturated $C_3N_4$ nanoparticle and exerts a significant effect on the electronic state composition for the CBM. The changes in the atomic structure and electronic properties described above indicate that the synergistic regulation of nanoparticle engineering, edge hydrogen saturation, and S doping provides an effective way to tune the atomic and electronic structures of bulk $\beta$-$C_3N_4$, which is of great significance for its application in photocatalysis and optoelectronic devices.

Unfortunately, for the case of S-doped hydrogenated NPs, we do not have the possibility of comparing with existing experimental data. On the contrary, in the case of g-$C_3N_4$, several experimental[32−35] and computational studies[77−79] exist, indicating new states in the band gap and/or a reduction of the band gap (exp. 2.55 eV).[35]

For the monolayer triazine-based g-$C_3N_4$ (Figure 7), the structures with different S doping sites replacing N atoms (Figure 10) clearly demonstrate the significant changes in the atomic and electronic structure caused by S doping. In the undoped case, g-$C_3N_4$ maintains the triazine unit periodic framework, with uniform C−N bond distribution, and its DOS exhibits typical semiconductor characteristics (Figure S2). However, for S-doped cases, different doping sites (S1 and S2) lead to significantly different structural reconstructions and DOS distributions.

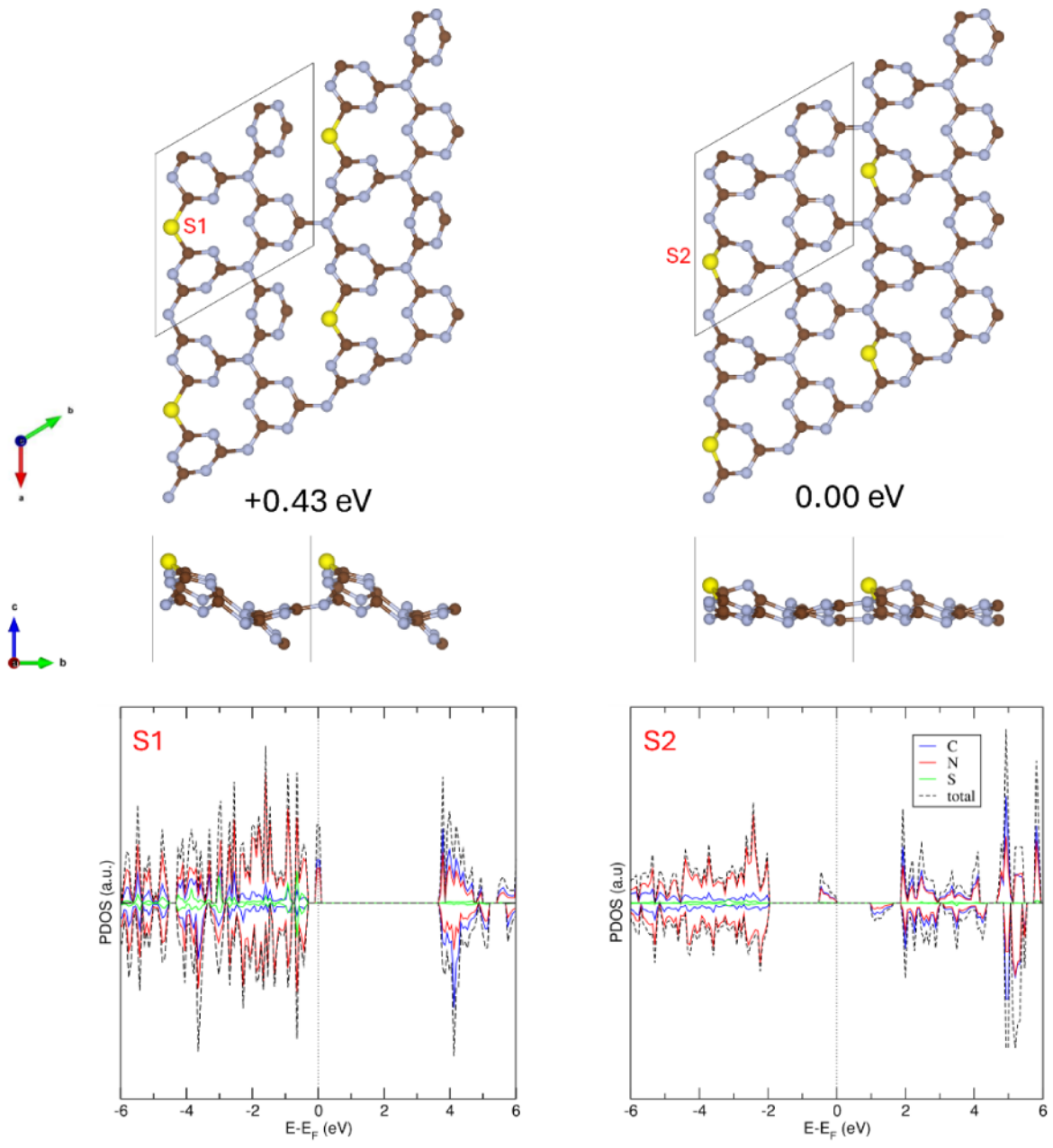

**Figure 10.** Atomic structure and DOS of monolayer triazine-based g-$C_3N_4$ with different sulfur (S) doping positions calculated by the HSE06+D3 method.

The energy of the S2 configuration is 0.43 eV per cell lower than that of S1, exhibiting significantly superior thermodynamic stability. This difference stems from the inherent bonding characteristics of the S atom, with the most stable chemical environment typically corresponding to forming two covalent bonds with surrounding atoms to satisfy the octet rule.

In the most stable S2 configuration, the S atom is coordinated to two surrounding C atoms. Because of the larger atomic radius of S compared to N, a noticeable bulge forms in the S2-doped system in the direction perpendicular to the ab plane (Figure 10). However, this geometric distortion is mainly confined to a local area and does not substantially disrupt the periodic interconnected frameworks of the surrounding triazine units. Therefore, the system can effectively release the local strain introduced by the heteroatom while maintaining the integrity of the basic $C_3N_3$ structural units, thus significantly reducing the structural distortion energy and obtaining the most stable configuration.

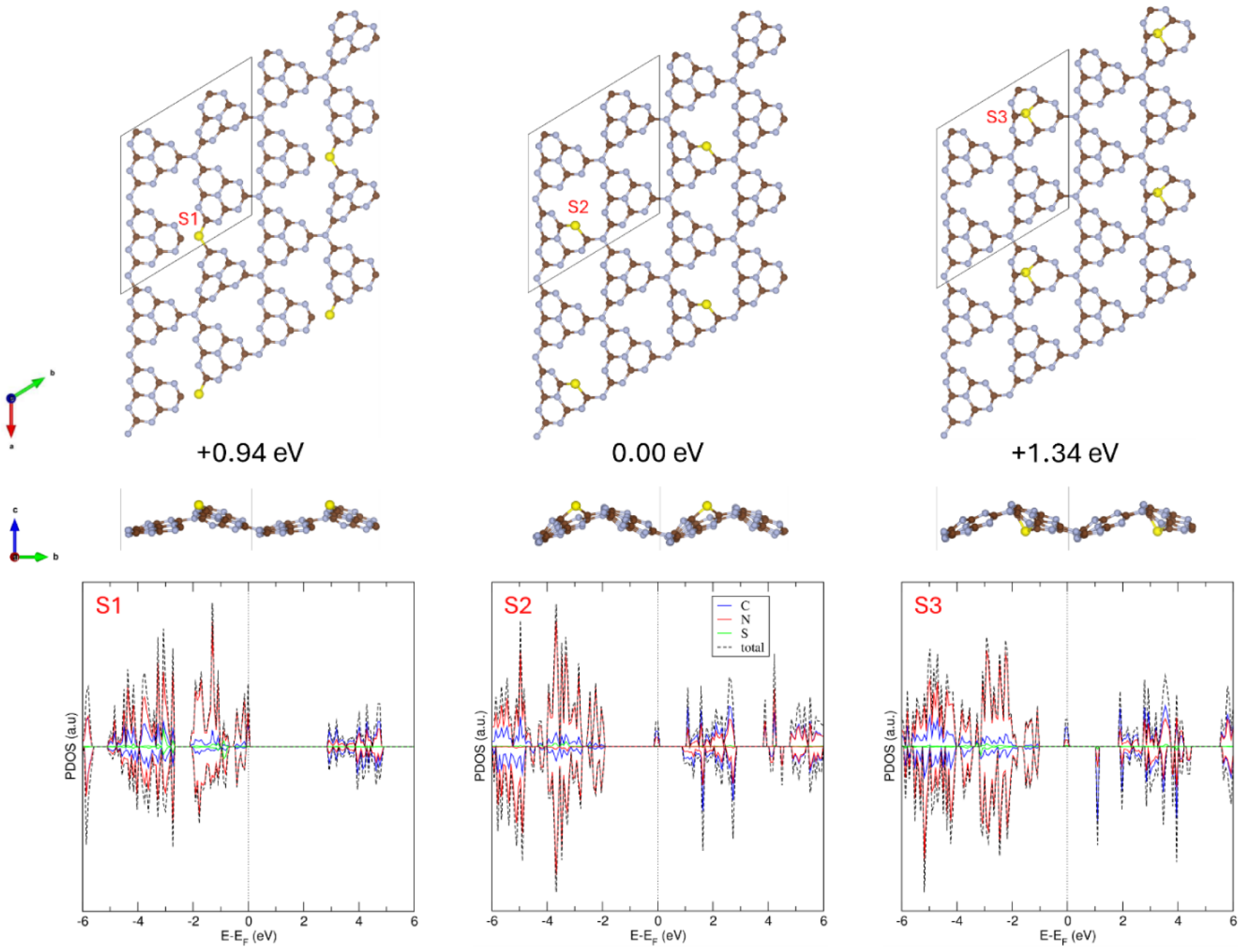

**Figure 11.** Atomic structure and DOS of monolayer heptazine-based g-$C_3N_4$ with different sulfur doping positions calculated by the HSE06+D3 method.

In contrast, in the S1 configuration (Figure 10), although the S atom is still coordinated with two surrounding C atoms, it breaks the bond with the third surrounding C atom, directly disrupting the originally periodically interconnected $C_3N_3$ triazine units. This bond breaking not only introduces significant local defects but also induces a chain reaction in the framework structure. Figure 10 shows that the S1 configuration is no longer limited to local geometric reconstruction but has undergone an integrated and significant structural deformation, indicating that the system needs to compensate for the chemical mismatch caused by doping through large-scale framework distortion. This integrated structural distortion significantly increases the strain energy, revealing why the S1 configuration is extremely unstable compared with the S2 configuration.

The introduction of one substitutional S-doping atom replacing one N atom led to an odd total number of electrons and, therefore, opened the possibility for a spin-polarized calculation where one unpaired electron is present in the system. Such solutions are always more stable than the nonspin-polarized ones (>0.4 eV). The extra unpaired electron in S1 is almost fully delocalized on the C with a dangling bond (0.9 e; see Figure S4), whereas in the case of S2, it is delocalized on three C atoms of the S-containing triazine unit (0.4 and 0.2 e on the two C atoms directly bonded to S and 0.3 on the C atom on the other side of the ring; see Figure S4).

The DOS plots in Figure 10 show that for S-doped triazine-based g-$C_3N_4$, S doping introduces new electronic states crossing the Fermi level in both the S1 and S2 configurations, but no true semiconductor-metal transition occurs in the system, in agreement with a previous study by Wang and Labat using HSE06-D2.[78] Although the DOS peaks appear at the Fermi level, they are narrow and isolated peaks, which are not mixed with either the valence or conduction bands, and hence are the typical doping-induced intermediate states. Furthermore, the DOS plots show that the Fermi state is primarily contributed by C atomic orbitals, followed by N atomic orbitals, while the contribution from S atomic orbitals is relatively small, clearly proved by the spin plots in Figure S4. This suggests that S atoms do not directly dominate the electronic states but indirectly influence the electronic structure by perturbing the C−N conjugated framework.

In summary, S doping does not significantly change the gap between the VBM and CBM of triazine-based g-$C_3N_4$ (about 3.8−3.9 eV; see Figure 10 and Figure S2). Instead, it introduces a doping-induced intermediate state, which enables photoexcitation to proceed through a two-step process of VBM → intermediate state → CBM, thereby significantly extending the optical absorption range.

Compared to monolayer heptazine-based g-$C_3N_4$ (Figure 8), the structures with different S-doping sites replacing N atoms (Figure 11) also exhibit significant atomic and electronic structure reconstructions. Undoped monolayer heptazine-based g-$C_3N_4$ has a periodic framework structure composed of $C_6N_7$ units (Figure 8). Its DOS spectrum shows no states crossing the Fermi level, exhibiting a clear semiconductor characteristic (Figure S3). The introduction of an S atom leads to significantly different local bonding environments and structural stability at different doping sites, namely, S1, S2, and S3 (Figure 11). The S2 configuration is the most stable, while the S1 and S3 configurations have energies about 0.94 and 1.34 eV per cell higher, respectively. This stability difference also stems from the inherent bonding characteristics of the S atom, whose most stable coordination environment is generally the formation of two covalent bonds with surrounding atoms. In the S2 configuration, the S atom is embedded in the heptazine framework in a two-coordinated manner with the surrounding C atoms. Its local geometric distortion is primarily released through vertical protrusions without disrupting the overall connectivity between $C_6N_7$ units, thus effectively lowering the structural distortion energy while maintaining the integrity of the conjugated $C_6N_7$ network. In contrast, the S1 and S3 configurations exhibit more pronounced structural distortions. In the S1 configuration, one S−C bond is broken, causing large structural distortions, not only the vertical undulations of the $C_6N_7$ units but also a more significant deviation from planarity. In the S3 configuration, although the S atom is bonded to three C atoms, the S−C bond lengths are asymmetric. The above analysis indicates that for S-doped heptazine-based g-$C_3N_4$, structural stability depends not only on the coordination environment but also on the distortion of the conjugated framework by the doping sites.

Also, in the case of heptazine, the introduction of one substitutional S-doping atom replacing one N atom led to an odd total number of electrons and, therefore, opened the possibility for a spin-polarized calculation where one unpaired electron is present in the system. Such solutions are always more stable than nonspin-polarized ones (greater than 0.4 eV). The extra unpaired electron in S1 is almost fully delocalized on the C with a dangling bond (0.9 e, see Figure S5), whereas in the case of S2 and S3, it is delocalized on more C atoms of the S-containing heptazine unit (three C atoms for S2 with 0.3, 0.4, and 0.2 and six C atoms for S3 with about 0.2 e each; see Figure S5).

The DOS plots (Figure 11) further indicate that S doping significantly modulates the electronic structure of monolayer heptazine-based g-$C_3N_4$, in good agreement with previous work by Wang et al.,[79] who used both PBE-D2 and HSE06 to investigate similar monolayers, and by Wang and Labat,[78] who investigated both free-standing and $TiO_2$ supported monolayers with HSE06-D2. Similar to the S-doped triazine-based system, the S1, S2, and S3 configurations all introduce new DOS peaks (doping-induced intermediate states) near the Fermi level. These near-Fermi states are also mainly contributed by C atomic orbitals, followed by N atomic orbitals, while the contribution of S atomic orbitals is relatively small, indicating that the S atom mainly indirectly regulates the electronic structure by perturbing the C−N conjugated network, proved by the spin plots in Figure S5.

A comparison of S-doped heptazine-based and triazine-based g-$C_3N_4$ reveals a high degree of similarity in the atomic and electronic structure regulation through S doping. The most stable configurations in both cases correspond to S atoms being embedded in the framework in a two-coordinated manner, releasing local strain without disrupting the periodic connections of the basic structural units. Heptazine-based g-$C_3N_4$ exhibits higher sensitivity to doping sites, with more significant differences in stability and DOS peaks between different doping configurations. This indicates that in heptazine-based g-$C_3N_4$ with its larger-scale conjugated framework, the rational selection of S doping sites is crucial for regulating the atomic and electronic structure. Overall, the S doping sites play a decisive role in the stability and electronic properties of g-$C_3N_4$, and the intrinsic bonding characteristic of the S atom preferring to form two covalent bonds is the fundamental reason for the differences in stability and electronic structure between different doping configurations. This pattern is clearly demonstrated in both triazine-based and heptazine-based g-$C_3N_4$. Furthermore, for both triazine-based and heptazine-based g-$C_3N_4$, S doping does not lead to a semiconductor-metal transition in these systems. Instead, it introduced highly localized intermediate states that transform these systems into typical doped semiconductors. These intermediate states allow photoexcitation to proceed through a two-step process of VBM → intermediate state → CBM, thereby significantly enhancing visible light absorption. This reveals the mechanism behind the significant redshift of the optical absorption of g-$C_3N_4$ induced by S doping, as observed in experiments.[80]

## 4. CONCLUSIONS

In this systematic DFT study, we have compared the standard PBE functional with the hybrid HSE06, both without and with the Grimme correction for dispersion forces (D3), to assess their ability in reproducing structural and electronic properties of different bulk phases of $C_3N_4$, ranging from diamond-like to graphitic layered structures. We found that, with all methods considered, layer corrugation enhances structural stability. HSE06-D3 is observed to well reproduce available experimental data and more sophisticated $G_0W_0$ results. For this reason, it was also used to investigate the nature of bulk triplet excitons in terms of excitation energy, self-trapping energy, and spin localization/delocalization. In the $\beta$-$C_3N_4$, the self-trapped triplet exciton is very localized, with the formation of two radicals at neighboring C and N atoms, whereas in g-$C_3N_4$, the spin is found to be delocalized within one building block unit (either triazine or heptazine in the two fully polymerized graphitic phases considered in this work), as a consequence of the interruption of the $\pi$-conjugation at the N linking nodes. Furthermore, we have considered the effect of nanostructuring on structural and electronic properties in terms of nanoparticle formation for $\beta$-$C_3N_4$ and single or multilayer exfoliation in the case of g-$C_3N_4$ structures. Finally, the modifications in the atomic and electronic structure as a consequence of extrinsic doping with nonmetal atoms (sulfur) were evaluated. At the surface of $\beta$-$C_3N_4$ nanoparticles, the presence of S atoms introduces new antibonding states at the bottom of the conduction band, whereas when substituting N atoms in the layers of g-$C_3N_4$, it causes the introduction of new localized states deep inside the band gap.

## ■ ASSOCIATED CONTENT

### Data Availability Statement

Data supporting the conclusions of this study are included in the paper and Supporting Information and are available from the corresponding author upon reasonable request.

### Supporting Information

The Supporting Information is available free of charge at https://pubs.acs.org/doi/10.1021/acs.jpcc.6c02390.

DOS of 2D triazine- and heptazine-based g-$C_3N_4$, spin plots of monolayer triazine- and heptazine-based g-$C_3N_4$ with different sulfur doping positions, and a table reporting literature data on flat/corrugated structures (PDF)

Structural data (ZIP)

## ■ AUTHOR INFORMATION

### Corresponding Author

**Cristiana Di Valentin** − *Department of Materials Science, University of Milano-Bicocca, Milano 20125, Italy;* orcid.org/0000-0003-4163-8062; Email: cristiana.divalentin@unimib.it

### Authors

**Da Chen** − *Department of Materials Science, University of Milano-Bicocca, Milano 20125, Italy*

**Pietro Andreozzi** − *Department of Materials Science, University of Milano-Bicocca, Milano 20125, Italy*

**Giulia Frigerio** − *Department of Materials Science, University of Milano-Bicocca, Milano 20125, Italy;* orcid.org/0000-0003-4517-6432

**Daniele Perilli** − *Department of Materials Science, University of Milano-Bicocca, Milano 20125, Italy;* orcid.org/0000-0002-3082-3986

**Paulo Siani** − *Department of Materials Science, University of Milano-Bicocca, Milano 20125, Italy;* orcid.org/0000-0002-1930-4579

Complete contact information is available at: https://pubs.acs.org/10.1021/acs.jpcc.6c02390

### Author Contributions

#D.C. and P.A. equally contributed to this work.

### Notes

The authors declare no competing financial interest.

## ■ ACKNOWLEDGMENTS

The authors are grateful to Dr. Chiara Daldossi for the technical help with the nanoparticle calculations. D.P. and C.D.V. acknowledge funding from the European Union-NextGenerationEU through the Italian Ministry of University and Research under PNRR-M4C2I1.4 ICSC-Centro Nazionale di Ricerca in High-Performance Computing, Big Data, and Quantum Computing (grant no. CN00000013).

## ■ REFERENCES